\documentclass[letterpaper,english,aps,letter,superscriptaddress,nofootinbib,floatfix,pre,preprint]{revtex4}
\usepackage[T1]{fontenc}
\usepackage[latin9]{inputenc}
\usepackage{amsbsy}
\usepackage{graphicx}
\usepackage{amssymb}

\makeatletter

\providecommand{\tabularnewline}{\\}

\usepackage{float}
\floatstyle{ruled}
\newfloat{algorithm}{tbp}{loa}
\floatname{algorithm}{Algorithm}
\renewcommand{\citet}[1]{\cite{#1}}

\usepackage{ae,aecompl}

\newenvironment{Algorithm}[1] {
  \refstepcounter{algorithm} 
  \vspace{1ex} \hrule \vspace{1ex} 
  \center{\textbf{Algorithm \thealgorithm : }#1}
  \vspace{1ex} \hrule \vspace{1ex}
  \small
} 
{ \normalsize \vspace{1ex} \hrule \vspace{1ex} }

\makeatother

\usepackage{babel}

\begin{document}

\title{A First-Passage Kinetic Monte Carlo Algorithm for Complex Diffusion-Reaction
Systems}

\author{Aleksandar Donev}

\affiliation{Lawrence Livermore National Laboratory, Livermore, California 94551,
USA}

\author{Vasily V. Bulatov}

\affiliation{Lawrence Livermore National Laboratory, Livermore, California 94551,
USA}

\author{Tomas Oppelstrup}

\affiliation{Lawrence Livermore National Laboratory, Livermore, California 94551,
USA}

\affiliation{Royal Institute of Technology (KTH), Stockholm S-10044, Sweden}

\author{George H. Gilmer}

\affiliation{Lawrence Livermore National Laboratory, Livermore, California 94551,
USA}

\author{Babak Sadigh}

\affiliation{Lawrence Livermore National Laboratory, Livermore, California 94551,
USA}

\author{Malvin H. Kalos}

\affiliation{Lawrence Livermore National Laboratory, Livermore, California 94551,
USA}

\begin{abstract}
We develop an asynchronous event-driven First-Passage Kinetic Monte
Carlo (FPKMC) algorithm for continuous time and space systems involving
multiple diffusing and reacting species of spherical particles in
two and three dimensions. The FPKMC algorithm presented here is based
on the method introduced in {[}\emph{Phys. Rev. Lett.}, 97:230602,
2006] and is implemented in a robust and flexible framework. Unlike
standard KMC algorithms such as the $n$-fold algorithm, FPKMC is
most efficient at low densities where it replaces the many small hops
needed for reactants to find each other with large first-passage hops
sampled from exact time-dependent Green's functions, \emph{without}
sacrificing accuracy. We describe in detail the key components of
the algorithm, including the event-loop and the sampling of first-passage
probability distributions, and demonstrate the accuracy of the new
method. We apply the FPKMC algorithm to the challenging problem of
simulation of long-term irradiation of metals, relevant to the performance
and aging of nuclear materials in current and future nuclear power
plants. The problem of radiation damage spans many decades of time-scales,
from picosecond spikes caused by primary cascades, to years of slow
damage annealing and microstructure evolution. Our implementation
of the FPKMC algorithm has been able to simulate the irradiation of
a metal sample for durations that are orders of magnitude longer than
any previous simulations using the standard Object KMC or more recent
asynchronous algorithms.
\end{abstract}
\maketitle
\newcommand{\Cross}[1]{\left|\mathbf{#1}\right|_{\times}}
\newcommand{\CrossL}[1]{\left|\mathbf{#1}\right|_{\times}^{L}}
\newcommand{\CrossR}[1]{\left|\mathbf{#1}\right|_{\times}^{R}}
\newcommand{\CrossS}[1]{\left|\mathbf{#1}\right|_{\boxtimes}}

\newcommand{\D}[1]{\Delta#1}
\newcommand{\V}[1]{\boldsymbol{#1}}
\newcommand{\M}[1]{\boldsymbol{#1}}

\newcommand{\grad}{\boldsymbol{\nabla}}
\newcommand{\eij}{\left\{  i,j\right\}  }
\newcommand{\celsius}{^{\circ}C}

\section{Introduction}

Models involving random walks are widely applied in science, engineering,
medicine and finance. Of particular interest are diffusion-reaction
systems in which multiple walkers walk simultaneously and independently
and some significant events take place when two or more walkers find
each other in space, or collide. Examples include formation and growth
of aggregates of colloidal particles in suspensions, kinetics of aerosols
in meteorology, diffusive phase transformations in solids \citet{Coarsening_AtomisticMeanField},
surface diffusion during crystal growth from vapor \citet{KMC_SurfaceReactions,LG_KMC_EpitaxialGrowth},
defect evolution in solids \citet{BIGMAC,LAKIMOCA_OKMC}, multi-particle
diffusion-limited aggregation in physics, diffusion-controlled reactions
in chemistry and biochemistry \citet{ParticleBased_CellModeling,GFRD_KMC,DiffusionReaction_Plimpton},
population dynamics, quantum physics \citet{KLV_PRE}, and risk assessment
and pricing models in finance, to name a few. Numerical simulations
of such processes often utilize various flavors of the Monte Carlo
method.

A novel diffusion Kinetic Monte Carlo (KMC) algorithm for simulating
diffusion-reaction systems in one, two, and three dimensions was first
proposed in Ref. \citet{FPKMC_PRL} and described in detail in Ref.
\citet{FPKMC1}. The algorithm cures the notorious inefficiency of
standard KMC algorithms \citet{BIGMAC} at low densities of the reacting
and diffusing particles: when diffusion is simulated via a sequence
of small hops, many such hops are needed to bring reactants together
and the standard algorithms are unable to progress over sufficiently
long time scales. The essential idea behind the new First Passage
Kinetic Monte Carlo (FPKMC) algorithm is to replace the long sequences
of small hops with large super-hops sampled from an exact Green's
function derived for a simpler auxiliary problem in which the diffusing
particles are spatially isolated (protected) from each other and thus
diffuse independently. The resulting algorithm is not only fast even
at low densities, but it is also exact (modulo numerical precision)
and does not, in principle, introduce approximations employed in other
accelerated algorithms, such as discretizing continuum diffusion into
a sequence of hops \citet{BIGMAC} or neglecting less-probable reactions
\citet{JERK,GFRD_KMC}.

Reference \citet{FPKMC1} focuses on the basic ideas behind the new
FPKMC method and illustrates its application in the context of simple
models in which identical or nearly identical cube-shaped particles
randomly walk on a lattice or in the continuum. Here we extend the
method to more general and complex diffusion-reaction systems, such
as multiple species with different sizes and diffusion coefficients,
competing reaction mechanisms, e.g. particle conversion, death and
insertion, absorbing boundary conditions, focusing on the case of
continuum isotropic diffusion without advection. We give a general
and formal presentation of the FPKMC method as an event-driven asynchronous
algorithm implemented for the case of additive hard spheres in two
and three dimensions. We also present algorithmic details, including
pseudo-codes, for a flexible yet efficient implementation of the FPKMC
method capable of handling a variety of problems of interest, including
radiation damage in metals \citet{JERK,JERK_CD_Barbu,LAKIMOCA_OKMC},
dopant implantation \citet{BIGMAC}, surface reactions \citet{KMC_SurfaceReactions,LG_KMC_EpitaxialGrowth},
coarsening \citet{Coarsening_AtomisticMeanField}, (bio)chemical reaction
networks \citet{ParticleBased_CellModeling,GFRD_KMC,DiffusionReaction_Plimpton},
and others. We validate the new algorithm by comparing our simulation
results against the Object KMC BIGMAC code \citet{BIGMAC} for several
non-trivial test problems. We then apply the new algorithm to simulations
of irradiation damage accumulation in iron and validate the new method
by comparing to results obtained using the existing object KMC code
LAKIMOCA \citet{LAKIMOCA_OKMC}. Finally, we demonstrate that the
new FPKMC algorithm allows extending the time horizon of radiation
damage simulations well beyond current computational limits and to
reach, for the first time, the long time-scales of material life in
a nuclear reactor (section \ref{SectionReactor}). 

In the remainder of this section we specify our diffusion-reaction
model and briefly discuss asynchronous event-driven algorithms. In
Section \ref{SectionFPKMC} we describe the core of the FPKMC algorithm,
namely, the use of exact time-dependent Green's functions for a suitably-defined
separable sub-problem. The new method is rather general and extends
to a variety of problems where diffusion plays a role, including also
discrete (lattice) systems and more general types of diffusion. In
Section \ref{SectionPairPropagators} we discuss numerical evaluation
of the time-dependent Green's functions for the case of hard spheres.
Section \ref{SectionAlgorithmic} gives further details, including
detailed pseudocodes for key components of the FPKMC algorithm. In
Section \ref{SectionResults} we present numerical validation of the
algorithm along with some performance figures, and finally, in Section
\ref{SectionConclusions} we offer a few concluding remarks.

\subsection{\label{SectionModel}Model Representation}

Consider a simulation of the time evolution of a collection of $N$
diffusing reactive particles in $d$-dimensions. For simplicity, we
will focus on the case of hard spheres of fixed radius diffusing in
a homogeneous medium. In the absence of reactions or surfaces and
assuming a particle started in some specific point $\V{r}_{0}$ at
time $0$, the probability $c(\V{r},t)$ of finding the same particle
in position $\V{r}$ at time $t$ is the solution to the time-dependent
diffusion equation \begin{equation}
\partial_{t}c=D\grad^{2}c\mbox{ and }c(\V{r},0)=\delta(\V{r}-\V{r}_{0}),\label{eq:basic_diffusion}\end{equation}
where $D$ is the particle diffusion coefficient. 

At any point in time, the state of the system is characterized by
its \emph{configuration} $\V{Q}=\left(\V{q}_{1},\ldots,\V{q}_{N}\right)$.
The number of particles $N$ may itself vary with time. Each particle
$i$ can possess an arbitrary number of attributes $\V{a}_{i}$ in
addition to the position of its centroid $\V{r}_{i}$, $\V{q}_{i}=\left(\V{r}_{i},\V{a}_{i}\right)$.
These attributes include a species $1\leq s_{i}\leq N_{s}$, a radius
$R_{i}$, a diffusion coefficient $D_{i}$, as well as other problem-specific
attributes such as charge, mass, etc. Some attributes may be shared
by all particles of a given species, for example, all particles of
a given species may have the same diffusion coefficient.

The symmetric \emph{reaction table} $\mathcal{R}_{\alpha\beta}$ of
size $N_{s}(N_{s}+1)/2$ specifies the type of \emph{two-particle
reaction} that occurs when particles $A$ and $B$ of species $\alpha$
and $\beta$ collide. In particular, $\mathcal{R}_{\alpha\beta}$
can specify that particles of species $\alpha$ and $\beta$ do not
interact with each other. Examples of possible reactions are \emph{annihilation}
$A+B\rightarrow0$, \emph{chemical reaction} $A+B\rightarrow C$,
including the special case of \emph{absorption} $A+B\rightarrow A$,
\emph{coalescence} $A+B\rightarrow AB$, and \emph{reflection} $A+B\rightarrow A+B$.
The \emph{decay table} $\mathcal{D}_{k}^{\alpha}$ specifies an arbitrary
number $N_{d}^{\alpha}$ of possible \emph{single-particle reactions}
for particles of species $\alpha$, assumed to occur as a Poisson
process with rates $\Gamma_{k}^{\alpha}=(\tau_{k}^{\alpha})^{-1}$.
Examples of decay reactions are \emph{splitting} $A\rightarrow B+C$,
including the special case of \emph{emission} $A\rightarrow A+B$,
\emph{transmutation} $A\rightarrow B$, \emph{death} $A\rightarrow0$,
and \emph{jump} (kick) $A\rightarrow A$. The positions of any products
of the single- or two-particle reactions are assigned depending on
the positions and attributes of the reactants, sometimes with additional
random displacements. If hard-wall boundaries are present, particles
colliding with a \emph{hard wall} $k$ they may be absorbed or reflected
with certain pre-specified probabilities.

The \emph{insertion rates} $\mathcal{B}_{\alpha}$ specify the rate
of insertion (birth) for particles of species $\alpha$ per unit time
per unit volume. Typically particles are inserted randomly and uniformly
inside the simulation volume. Note that in some applications it can
be necessary to insert a whole collection of particles rather than
a single particle. For example, irradiation by heavy particles - ions
or neutrons - creates large displacement cascades that quickly anneal
into a whole collection of point defects and clusters.

\subsection{\label{SectionAED}Asynchronous Event-Driven Algorithms}

The First-Passage Kinetic Monte Carlo (FPKMC) algorithm belongs to
the category of \emph{asynchronous} event-driven (AED) algorithms
\citet{AED_Review}. The algorithm is similar to the well-known event-driven
molecular dynamics (EDMD) algorithm \citet{EventDriven_Alder}, with
the essential difference that in FPKMC particle dynamics is stochastic
rather than deterministic \citet{EventDriven_Colloids}. Just like
the hard-sphere Molecular Dynamics, FPKMC is an \emph{exact} algorithm
(within numerical precision) because the two-particle problem in FPKMC
can be solved exactly. By its exactness we mean that FPKMC samples
evolution trajectories of $N$ random walkers from the correct probability
distribution, as given by the exact solution of an appropriate Master
equation.

Event-driven algorithms evolve the state of the system by updating
it only when certain non-trivial \emph{events} occur, skipping the
time elapsed between such events as uninteresting, e.g., unchanged
or analytically solvable. In the asynchronous algorithms, there is
a global simulation time $t$, typically the time when the last processed
event occurred, and each particle $i$ is associated with a point
in time $t_{i}\leq t$, typically the last time it participated in
an event. This is to be contrasted to synchronous event-driven algorithms,
where all of the particles are at the same time $t$, such as the
$n$-fold (BKL) algorithm for performing kinetic (dynamic) Monte Carlo
simulations \citet{Ising_BKL}. The classical $n$-fold algorithm
hinges on the fact that the state of the system does not change between
the events, as is common in lattice models where particle positions
are discrete. For example, the atoms may vibrate around the lattice
sites and occasionally hop to nearby sites. In the model considered
here, however, the positions of the particles are continuous and continuously
changing even between events.

An asynchronous simulation progresses by processing the event at the
head of the queue, scheduling new events for any affected particles,
and then updating the event queue. The main types of events in FPKMC
and their scheduling and processing will be described in detail in
the next section. Each particle $i$ stores the time it was last updated
$t_{i}$ along with a prediction for its \emph{impending event} $\left(t_{i}^{e},p_{i},\nu_{i}\right)$,
specified via the predicted time of occurrence (timestamp) $t_{i}^{e}$,
the event \emph{partner} $p_{i}$, and the event \emph{qualifier}
(type of event) $\nu_{i}$. When it is clear what particle we are
referring to, we will omit the subscript $i$ for simplicity. If several
different events are possible for a given particle then the first
event scheduled to occur (i.e., the one with the smallest $t_{e}$)
is chosen. The event times for all particles and event times for any
\emph{external events} are stored in a priority queue (e.g., a heap)
called the \emph{event queue}. If the partner $p=j$ is another particle,
then event prediction for the partner particle $j$ is not stored
in the event queue to avoid sorting duplicate events with equal timestamps.

\section{\label{SectionFPKMC}The First-Passage Kinetic Monte Carlo Algorithm}

A detailed description of the FPKMC algorithm, including pseudo-codes,
is given in Section \ref{SectionAlgorithmic}. Here we only briefly
discuss the most important components of the algorithm. Although this
description is intended to be self-contained, the reader is referred
to Ref. \citet{FPKMC1} for a more intuitive introduction.

The essential idea behind the FPKMC algorithm is to break the $N$-body
problem into a collection of independent one-body or two-body (pair)
problems that can be solved analytically. This is achieved by protecting
each particle $i$ with a \emph{protective region} $\mathcal{P}_{i}$,
$\mathcal{C}_{i}\subseteq\mathcal{P}_{i}$, where the hard core of
particle $i$ is denoted by $\mathcal{C}_{i}$. An unprotected particle
has $\mathcal{P}_{i}\equiv\mathcal{C}_{i}$. In the case of hard spheres
\footnote{The cube-shaped particles and protections described in Ref. \citet{FPKMC1}
can be thought of as spheres but with an $L_{\infty}$ distance metric
function $\left\Vert \D{\V{r}}\right\Vert =\max_{1\leq k\leq d}\D{r}_{k}$,
instead of the $L_{2}$ Euclidean distance $\left\Vert \D{\V{r}}\right\Vert =\sqrt{\sum_{k=1}^{d}\D{r}_{k}^{2}}$
characteristic of spheres.%
}, $\mathcal{C}_{i}=\left\{ \V{r}\left|\mbox{ }\left\Vert \V{r}-\V{r}_{i}\right\Vert \leq R_{i}\right.\right\} $,
the protective regions themselves are spheres of radius $R_{i}^{(\mathcal{P})}>R_{i}$
centered at $\V{r}_{i}^{(\mathcal{P})}$. Thus, a spherical particle
of radius $R_{i}$ can be thought of as a \emph{point particle} contained
inside a protective sphere of radius $R_{i}^{(\mathcal{P})}-R_{i}$,
i.e. $\left\Vert \V{r}_{i}^{(\mathcal{P})}-\V{r}_{i}\right\Vert \leq R_{i}^{(\mathcal{P})}-R_{i}$.
In general it is not required that $\V{r}_{i}^{(\mathcal{P})}=\V{r}_{i}^{0}$,
however, making the protective sphere concentric with the particle
simplifies implementation.

\subsection{First-Passage Probability Densities}

If the protective region $\mathcal{P}_{i}$ of a particle $i$ is
disjoint from the protective regions of other particles, then the
diffusive motion of the particle is independent of the motion of other
particles, \emph{as long as the particle is still inside its protection}.
The motion of the particle inside its protection is a one-body diffusion
problem that can often be solved analytically. The FPKMC algorithm
entails sampling from the following two probability distribution functions
(PDFs) for a particle initially at $\V{r}_{0}=0$ at time $t=0$:

\begin{enumerate}
\item The \emph{first-passage probability distribution} $J_{1}(\tilde{t},\tilde{\V{r}})$
that the particle first leaves its protective region at time $\tilde{t}$,
when it is at position $\tilde{\V{r}}$. We call the time $\tilde{t}$
the \emph{exit time} and the position $\tilde{\V{r}}$ the \emph{exit
location}. For a point particle, $\tilde{\V{r}}$ is on the surface
of the protection, $\tilde{\V{r}}\in\partial\mathcal{P}_{i}$.
\item The conditional \emph{no-passage probability distribution} $c_{1}(\V{r};t)$
that the particle is at position $\V{r}\in\mathcal{P}_{i}\setminus\partial\mathcal{P}_{i}$
at a given time $t$, \emph{given} that it has not left its protection
by time $t$.
\end{enumerate}
These two probability distributions are the basic elements of the
theory of first-passage processes \citet{FirstPassage_Redner} and
are termed hereafter the first-passage and no-passage\emph{ propagators},
respectively. Single-particle propagators for spherical particles
inside spherical protection regions are discussed in Section \ref{SectionMonomers}.

\subsubsection{Pair Propagators}

In the FPKMC algorithm, the particles are protected by disjoint protective
regions allowing the use of single-particle propagators to evolve
the system. At some point in time, however, two particles $i$ and
$j$ will collide and thus cannot be protected with disjoint regions.
For efficient handling of collision events, nearly-colliding pairs
are \emph{associated} (partnered) and protected by a \emph{pair protection
region} $\mathcal{P}_{ij}$. We will focus on the case when $\mathcal{P}_{ij}=\mathcal{P}_{i}\cup\mathcal{P}_{j}$
consists of intersecting protections around each of the two particles,
$\mathcal{P}_{i}\cap\mathcal{P}_{j}\neq\emptyset$. If either one
of the particles leaves its protective region the pair \emph{disassociates}.

Note that, for the case of additive hard spheres that we consider
here, triple collisions never happen. Thus, it will always be possible
to protect two colliding particles $i$ and $j$ as a pair even if
there is a third particle $k$ nearby. As the simulation time approaches
the collision time, eventually $i$ and $j$ will be much closer to
each other than to $k$ and can thus be pair-protected with protective
region $\mathcal{P}_{ij}$ disjoint from $\mathcal{P}_{k}$. For non-additive
hard spheres or other types of collision rules pair protection may
not be sufficient and in such cases the approximate handling of interactions
discussed in Section \ref{SectionTimeDriven} will be useful. 

For an associated pair of particles, in general, the FPKMC algorithm
requires sampling from the following two distributions:

\begin{enumerate}
\item The first-passage probability distribution $J_{2}(\tilde{t},\tilde{\V{r}}_{i},\tilde{\V{r}}_{j})$
that at time $\tilde{t}$, when the particles are at positions $\tilde{\V{r}}_{i}$
and $\tilde{\V{r}}_{j}$ respectively, one of two particles of the
pair leaves its protection for the first time, or particles $i$ and
$j$ collide and react for the first time.
\item The conditional no-passage probability distribution $c_{2}(\V{r}_{i},\V{r}_{j};t)$
for the positions of the particles at a given time $t$, \emph{given}
that neither $i$ nor $j$ has left its protection, nor a collision
has occurred. 
\end{enumerate}
We term these two distributions \emph{two-particle }or\emph{ pair
propagators}. As discussed in part I of this series, the pair propagators
allow further factorization into two single-body propagators. The
first propagator is for the \emph{difference walker} $\V{r}_{ij}^{(D)}=\D{\V{r}}_{i}-\D{\V{r}}_{j}$,
where the condition $\left\Vert \V{r}_{ij}^{(D)}\right\Vert =R_{i}+R_{j}$
corresponds to a collision. The second propagator is for the \emph{center
walker} $\V{r}_{ij}^{(C)}=w_{i}\D{\V{r}}_{i}+w_{j}\D{\V{r}}_{j}$,
where $w_{i}$ and $w_{j}$ are appropriately-chosen weights. The
protections and the propagators for the difference and center walkers
are discussed in further detail in Section \ref{SectionPairs}.

\subsubsection{Hard-Wall Propagators}

In some situations periodic boundary conditions may not be appropriate
and instead hard-wall boundaries could be used along certain directions
of the simulation box. Furthermore, additional absorbing or reflecting
surfaces may need to be inserted in the simulation volume, for example,
to represent grain boundaries in a polycrystalline material. Particle
and pair protections should be disjoint from any hard-wall surfaces
or boundaries $\mathcal{W}$ in order to ensure that single-particle
and pair propagators can be used. However, some particle $i$ will
eventually collide with $\mathcal{W}$ and, thus, can not remain protected
from $\mathcal{W}$ forever. Therefore, particles near a wall $\mathcal{W}$
are associated (paired) with the wall itself and protected with a
\emph{hard-wall protection} $\mathcal{P}_{i}$ that is disjoint from
all other single-particle and pair protections but intersects wall
$\mathcal{W}$, $\mathcal{P}_{i}\cap\mathcal{W}=\partial\mathcal{P}_{i}^{\mathcal{W}}\neq\emptyset$.
For such particles, we will use of the following \emph{hard-wall propagators}:

\begin{enumerate}
\item The first-passage probability distribution $J_{HW}(\tilde{t},\tilde{\V{r}})$
that the particle exits its protective region or collides with the
hard wall at time $\tilde{t}$, when it is at position $\tilde{\V{r}}$.
\item The conditional no-passage probability distribution $c_{HW}(\V{r};t)$
that the particle is at position $\V{r}\in\mathcal{P}_{i}\setminus\partial\mathcal{P}_{i}$
at a given time $t$, \emph{given} that it has not left its protection
or collided with the wall by time $t$.
\end{enumerate}

\subsection{\label{SectionSummaryOfFPKMC}Summary of the FPKMC Algorithm}

As already discussed in Section \ref{SectionAED}, the FPKMC algorithm
processes a sequence of events in the increasing time order in an
event loop. Each event has a \emph{qualifier} (type of event) and
is associated with a \emph{primary} particle, which may have an event
\emph{partner} $p_{i}$. In the implementation of the algorithm described
here the particles can only have one partner, either another particle
forming a pair, or a hard wall. In general one can have multi-particle
events and treat a group of particles together. We discuss this modification
again in Section \ref{SectionTimeDriven} and will report on its implementation
details in future publications.

Following Refs. \citet{FPKMC_PRL,FPKMC1}, the core of the FPKMC algorithm
entails the following generic steps:

\begin{enumerate}
\item \label{ConstructInitialProtections}Construct initial protective regions
around all particles, either single-particle protections not overlapping
with other protections or boundaries, pair protections for pairs of
particles much closer to each other than to other particles, or hard-wall
protections for particles much closer to a boundary than to other
particles.
\item Sample an exit time for each particle or pair (in the case of pairs
this can mean a collision) from the corresponding first-passage probability
distribution, and make an event queue with all of the scheduled events. 
\item \label{ProcessNextExit}Using the event queue, find the shortest event
time and identify the corresponding particle(s). Sample the exit position(s)
from the corresponding first-passage probability distribution. If
a collision occurs, take appropriate action.
\item \label{ConstructNewProtections}Construct new protections for the
particles propagated in Step \ref{ProcessNextExit}. If necessary
to make more space available for protection of the propagated particle(s),
sample new locations for nearby particles from the corresponding no-passage
probability distribution, and re-protect those particles as well. 
\item Sample new event times for the particle(s) protected in step \ref{ConstructNewProtections},
and update the corresponding events in the event queue. Go back to
step \ref{ProcessNextExit}.
\end{enumerate}
Further details on the main event loop are given in Section \ref{SectionAlgorithmic}.

In FPKMC, the most important and frequent events are first-passage
propagations in which a particle or a pair reaches the boundary of
its protective region or collides. Such events are scheduled and processed
by sampling from appropriate first-passage distributions, and then
processing any reactions, as appropriate. For a given set of particle
starting positions, one can choose to build protections to maximize
the expectation value of the next simulated time increment. To achive
this, $N$ particles should be protected in such a way that the expectation
value for the minimal first-passage time is maximized:

\[
\max_{\mbox{Protections}}E\left[\min\{t_{1},t_{2},...,t_{N}\}\right].\]
Finding the optimal space partitioning that maximizes the above expectation
value is a non-trivial problem of constrained non-linear optimization.
It is especially difficult to solve in the course of an asynchronous
FPKMC simulation where computational cost of space re-partitioning
should be balanced with the resulting benefit to the time increment.
Finding the optimal space partitioning is further complicated in a
dynamic context, when FPKMC propagation events compete with other
stochastic processes such as particles creation, emission, destruction,
etc. In the current FPKMC impementation, we use a simpler optimality
condition in which the minimum of $N$ expected first-passage times
is maximized:

\[
\max_{\mbox{Protections}}\min\{E(t_{1}),E(t_{2}),...,E(t_{N})\}.\]
Obviously, the protections have to be as large as possible, with more
space given to particles with higher diffusion rates. To better quantify
this requirement, let us first consider just two particles and protect
them by two concentric non-overlapping protection domains $\Omega_{1}$
and $\Omega_{2}$. The expectation of the earliest of two exit times
is maximal when domains $\Omega_{1}$ and $\Omega_{2}$ touch each
other and their linear dimensions are proportional to the square roots
of two diffusion coefficients $\sqrt{D_{1}}$ and $\sqrt{D_{2}}$.
In this case the expected first passage times for both particles are
equal. Let us now use this maximal expected exit time as a measure
of distance between any two unprotected particles. If all particles
are unprotected, solving for the optimal protection amounts to finding,
using the so-defined distance metric, the nearest neighbors in the
whole system. If possible, one protects the nearest neighbors as a
pair, otherwise, they are protected individually by non-overlapping
(touching) protective regions. 

The above method, however, only optimizes protections over a \emph{single}
first-passage event (\emph{locally-optimal} protections) and does
not necessarily maximize the expected time per event over a \emph{long
sequence} of events (\emph{globally-optimal} protections). Specifically,
following a single event, an optimal re-partitioning of the space
into protective regions could be constructed, however, this would
require destroying multiple existing protections and thus cause multiple
propagations and new event predictions. Obviously, it is necessary
to find a reasonable trade-off between re-building too many protections
and giving all of the particles sufficient room to move over sufficient
distances compared to other particles.

Following the processing of an event, typically one or two particles
are unprotected. We try to protect those particles, always starting
from a chosen \emph{seed} particle, with locally-optimal protections,
without disturbing any other protected particles. Specifically, in
our algorithm, one first finds the largest possible protection for
the seed particle by performing a neighbor search over all nearby
protected or unprotected particles. If the \emph{limiting neighbor}
is protected, one can simply protect the particle with a protection
that touches the limiting protection %
\footnote{Other choices are possible here, for example, one may look at the
expected or scheduled first-passage time for the limiting neighbor.%
}. Otherwise, one recursively applies the same procedure to the unprotected
limiting neighbor. If two particles are both unprotected and are found
to be mutually limiting neighbors, we attempt to protect them as a
pair provided they are sufficiently close. Detailed pseudocode for
this procedure is given in Section \ref{SectionRecursiveProtection}.

If the size of the single protection of the seed particle is too small,
we make room for a larger protection by destroying third-party protections.
That is, if an unprotected particle gets too close to the protective
region of another particle or pair, that particle or pair is brought
to the current point in time using the appropriate no-passage propagator.
Similarly, if the seed particle is nearly colliding with another particle
but the two particles cannot be protected as a pair, we destroy any
other protections blocking pair protection. Following the destruction
of other protections, we repeat the process starting from protection
of the seed particle. Instead of keeping a list of all unprotected
particles we simply schedule immediate re-protection events (i.e.,
place them at the head of the event queue) for any particles whose
protections were destroyed.

Our implementation relies on a number of input tolerances and heuristics
to manage particle and pair protections. In particular, the following
decisions need to be made by the algorithm and can be controlled with
various input tolerances:

\begin{itemize}
\item When should a pair of unprotected nearest-neighbor particles be selected
for pair protection, beyond the necessary conditions for pair protection
discussed in Section \ref{SectionPairs}? The computationally-optimal
choice here depends on the relative cost of the single and pair propagations.
We favor single protections because of their lower cost and build
pair protections only when two particles get closer to each other
than some fraction of their size.
\item When is it justified to unprotect a protected particle or a protected
pair in order to make room for protecting an unprotected single particle
or a pair candidate? One may use a hard cutoff here, however, this
introduces an artificial length-scale and the algorithm looses some
of its ability to adopt to the variations in particle density (and
thus typical sizes of the protective regions). We prefer to always
destroy the most limiting protection following a first-passage propagation.
\item If two unprotected nearest-neighbor particles are not a pair candidate,
how should the available space be divided among their single protections?
As described above, we split the space proportionally to the square
root of the diffusion coefficients of the two particles.
\item Should the size of the protective region of a single particle be limited
even if there is more room available for its protection (note that
a large protection is more likely to block future protections of other
particles)? We control this through an input parameter that set the
time limit $\D{t}_{max}$ after which the protection is almost certainly
going to be destroyed before the first-passage event actually occurs,
and limit the size of $\mathcal{P}_{i}$ by $\sqrt{D_{i}\D{t}_{max}}$.
\item Once it is determined that two nearby particles can be protected as
a pair, how large should their pair protection be (of course, respecting
the minimal required size discussed in Section \ref{SectionPairs}
and the maximal size allowed by other neighbors)? We control this
through an input target range for the parameter $\alpha$ defined
in Section \ref{SectionPairs}.
\end{itemize}
Intuitively, the algorithm's performance will be optimized by giving
more protective space to the particle or a small subset of particles
that dominate(s) the event queue, up to the point when giving more
room to these \emph{fast particles} no longer increases the average
time interval between subsequent events (or actually makes efficiency
worse by squeezing the other \emph{slow particles} too much). In Section
\ref{SectionOptimization} we will describe a self-tuning strategy
we applied in several specific situations encountered in FPKMC simulations.
While some of our strategies are relatively general, understanding
of the problem at hand helps considerably in optimizing the algorithm
performance.

We conclude this section with a few additional notes on the components
of the FPKMC algorithm. Whenever an event creates new particles, a
check is made to determine whether the newly inserted particle overlaps
with any existing particles. If it does, appropriate reactions are
immediately processed and the overlap check is repeated until no overlap
is detected. Additionally, unlikely the mobile particles, the immobile
particles remain unprotected and have no partners, but other particles
can have them as partners. This is particularly useful when there
are large immobile particles (e.g. clusters of monomers) that are
surrounded by a dense pool of small mobile particles (e.g., monomers).
The mobile particles do not affect each other for as long as their
protections do not overlap, even if they share the same immobile cluster
as a partner. However, when an event changes the configuration of
the immobile cluster (e.g., it absorbs a monomer), all its partner
particles are brought to the current time and re-protected.

\section{\label{SectionPairPropagators}Single Particle and Pair Propagators}

In this section we describe first-passage and no-passage propagators
for spherical particles. The propagators for single particles are
similar to the ones described in Ref. \citet{FPKMC1} for cube-shaped
particles and are discussed here only briefly. However, the pair propagators
for two spherical particles are considerably different than those
for cubical particles and are presented in more detail.

\subsection{\label{SectionMonomers}Single Particle Propagators}

The FPKMC algorithm requires first-passage and no-passage propagators
for a single spherical particle $A$ of radius $R_{A}$ with diffusion
coefficient $D_{A}$, starting from an initial position $\V{r}_{A}^{0}$
at the center of a protective sphere $\mathcal{P}_{A}$ of radius
$R_{A}^{\mathcal{P}}>R_{A}$ concentric with the particle. This reduces
to finding the propagators for a \emph{point }Brownian particle starting
at time $t=0$ from the center of sphere with radius $R_{P}^{\mathcal{P}}=R_{A}^{\mathcal{P}}-R_{A}$.
Due to the full rotational symmetry, the first-passage PDF $J_{1}(\tilde{t})$
is a function of time only, and the exit location on the protective
sphere $\mathcal{P}_{A}$ can be sampled from a uniform distribution.
Similarly, the no-passage PDF $c_{1}(r;t)$ becomes a function of
time and radial distance only, as the actual position $\V{r}$ can
be sampled from a uniform distribution on the sphere of radius $r$.

Expressing the distances and time in reduced units, the propagators
can be obtained by solving the diffusion problem for a point Brownian
particle with diffusion coefficient $D=1$ inside a sphere of unit
radius $R_{P}^{P}=1$, \[
\frac{\partial\left[rc(r,t)\right]}{\partial t}=\frac{\partial^{2}\left[rc(r,t)\right]}{\partial r^{2}}\]
with the boundary condition $c(1,t)=0$ and the initial condition
$c(r,0)=\delta(r)$. The probability to survive up to time $t$, i.e.
the \emph{survival probability}, is \[
S(t)=\int_{r=0}^{1}4\pi r^{2}c(r,t)dr.\]
The probability distribution for the exit time, i.e. the \emph{exit
probability}, is \begin{equation}
p(t>0)=-\frac{\partial S(t)}{\partial t},\label{eq:exit_probability}\end{equation}
and is used to sample the exit time $\tilde{t}$. The no-passage probability
distribution is \begin{equation}
c_{1}(r<1;t)=\frac{c(r,t)}{S(t)}.\label{eq:NP_propagator}\end{equation}
Series solutions for the various probability distribution functions
for the case of a point walker in a cube-shaped region were presented
in Ref. \citet{FPKMC1}. For the case of spherical particles inside
spherical protection regions we focus on two useful series expansions
of the solution, one that converges quickly at short times ($t\lesssim1/4$),

\begin{equation}
c(r,t)=(4\pi t)^{-\frac{3}{2}}\sum_{m=-\infty}^{\infty}(1+\frac{2m}{r})\exp\left[-\frac{(r+2m)^{2}}{4t}\right]\label{c_mono_short}\end{equation}
and another that converges quickly at long times ($t\gtrsim1/\pi^{2}$),

\begin{equation}
c(r,t)=\frac{1}{2r}\sum_{m=1}^{\infty}m\sin(m\pi r)e^{-m^{2}\pi^{2}t}.\label{c_mono_long}\end{equation}

The first-passage and no-passage distributions can be sampled numerically
in various ways. We use rejection sampling as proposed in Ref. \citet{FPKMC1}
and detailed in Appendix \ref{AppendixMonomer}. However, we emphasize
that rejection sampling is not the only nor necessarily the most efficient
method for sampling the required distributions, especially if one
is willing to accept some error (which can be controlled). For example,
direct tabulation and the use of lookup tables may be simpler and
more efficient in certain cases, especially when there is high symmetry
in the shape of the particles and protective domains. Such implementation
details are, however, orthogonal to the development of the event-driven
algorithm and can be improved upon separately.

\subsection{\label{SectionPairs}Pair Propagators}

In this section we discuss particle protection and propagators used
to enable collisions of pairs of particles. Consider two spherical
Brownian particles $A$ and $B$ of radii $R_{A}$ and $R_{B}$ and
diffusion coefficients $D_{A}$ and $D_{B}$. Each of the particles
is protected by a sphere of radii $R_{A/B}^{\mathcal{P}}>R_{A/B}$
concentric with the initial particle positions $\V{r}_{A/B}^{0}$.
When two protective spheres overlap, i.e., $R_{A}^{\mathcal{P}}+R_{B}^{\mathcal{P}}>r_{AB}=\left\Vert \V{r}_{A}-\V{r}_{B}\right\Vert $,
the particles can collide while diffusing within their individual
protections. Sampling of particle collisions is enabled by transforming
the two-particle diffusion problem into two single-particle problems,
one for the \emph{difference walker} $\V{r}_{D}=\V{r}_{A}-\V{r}_{B}$
and the other for the \emph{center walker} $\V{r}_{C}=w_{A}\V{r}_{A}+w_{B}\V{r}_{B}$.
A collision occurs when the difference walker reaches the collision
radius $r_{D}=\left\Vert \V{r}_{A}-\V{r}_{B}\right\Vert =R_{A}+R_{B}$.
It can be shown that, with the choice $w_{A}D_{A}=w_{B}D_{B}$, the
cross term $\frac{\partial^{2}}{\partial\V{r}_{C}\partial\V{r}_{D}}$
in the Laplacian operator vanishes which means that the six-dimensional
PDF for the pair factorizes into a product of two three-dimensional
PDFs, one for each walker. The resulting center and difference walkers
diffuse independently with diffusion coefficients $D_{D}=D_{A}+D_{B}$
and $D_{C}=w_{A}^{2}D_{A}+w_{B}^{2}D_{A}$. 

There is considerable freedom for choosing weights $w_{A}$ and $w_{B}$
and protections for the center and difference walkers. To simplify
the implementation, the center and difference walkers are each protected
by spheres of radii $R_{D}^{\mathcal{P}}$ and $R_{C}^{\mathcal{P}}$
centered around their initial positions. It turns out that choosing
$R_{D}^{\mathcal{P}}=R_{C}^{\mathcal{P}}$ maximizes the use of space
within the overlapping protective spheres of the two original particles.
Furthermore, by setting $D_{C}=D_{D}$ the expectation values for
the exit times for the difference and center walkers become similar
and nearly maximal. This leads to the following optimal choice for
the coordinate transformation\begin{eqnarray*}
\V{r}_{D} & = & \V{r}_{A}-\V{r}_{B}\\
\V{r}_{C} & = & \sqrt{\frac{D_{B}}{D_{A}}}\V{r}_{A}+\sqrt{\frac{D_{A}}{D_{B}}}\V{r}_{B}.\end{eqnarray*}
Finally, the condition that a collision should be possible requires
that $R_{C}^{\mathcal{P}}>\delta$, where $\delta=r_{AB}-(R_{A}^{\mathcal{P}}+R_{B}^{\mathcal{P}})$
is the initial inter-particle gap. We take\[
R_{C}^{\mathcal{P}}=R_{D}^{\mathcal{P}}=(2+\alpha)\delta,\]
where $\alpha\geq-1$ is a parameter in the algorithm. We find that
$\alpha=1$ or $R_{C}^{\mathcal{P}}=2\delta$ is a reasonable choice
but, when there is more room available, one can increase $\alpha$
up to a maximal value $\alpha_{max}$ specified in the code. It can
be shown that, as the difference and center walkers propagate inside
their own protections, each of the two original particles remains
within a sphere of radius\[
R_{A/B}^{\mathcal{P}}=R_{A/B}+(2+\alpha)\delta\sqrt{D_{A/B}}\frac{\sqrt{D_{A}}+\sqrt{D_{B}}}{D_{A}+D_{B}}.\]
This defines the size of the protective region around each original
particle that is minimally necessary to allow pair protection, i.e.,
the minimal distance to the next-nearest neighbors of particles $A$
and $B$ which allows for pair protection.

\subsubsection{The Difference and Center Propagators}

With the above coordinate transformation, the first-passage problem
for the pair of original particles separates into two independent
first-passage problems for the center walker and the difference walker.
The overall first-passage time $\tilde{t}$ is the smaller of the
two first-passage times. Thus, the possible first-passage events for
the pair are:

\begin{description}
\item [{Collision}] when the difference walker reaches the surface $r_{D}=\left\Vert \V{r}_{A}-\V{r}_{B}\right\Vert =R_{A}+R_{B}$. 
\item [{Dissolution}] when the difference walker reaches the surface $r_{D}=R_{D}^{\mathcal{P}}$.
The center walker has not yet left $\mathcal{P}_{C}$ and can be updated
using the single-point no-passage propagator $c_{1}(\D{\V{r}_{C}};\tilde{t})$.
\item [{Displacement}] when the center walker reaches the surface $r_{C}=R_{C}^{\mathcal{P}}$.
The difference walker has not yet left $\mathcal{P}_{D}$ and can
be updated using a special single-point no-passage propagator $c_{D}(\D{\V{r}_{D}};\tilde{t})$
(see the next section).
\end{description}
Because the center walker is restricted inside a spherical protection
$\mathcal{P}_{C}$ of radius $R_{C}^{\mathcal{P}}$, the same FP and
NP propagators described earlier in Section \ref{SectionMonomers}for
single particles can be used for this walker.

\begin{figure}
\noindent \begin{centering}
\includegraphics[width=0.5\textwidth]{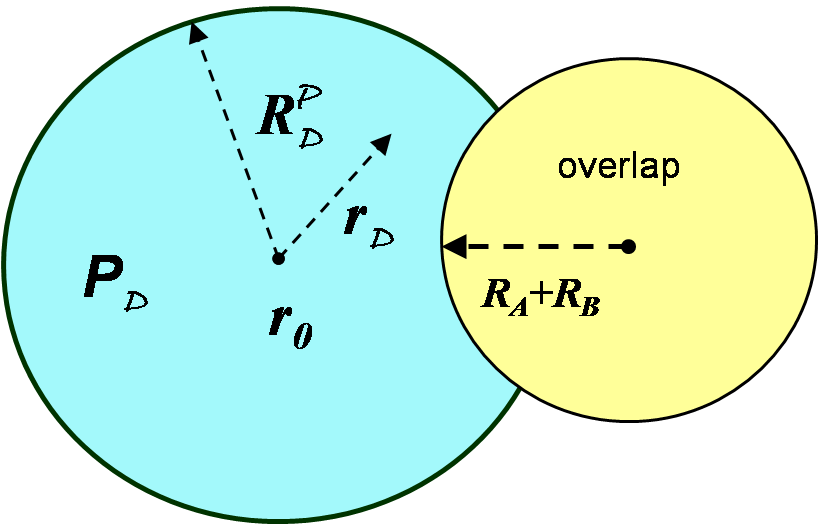}
\par\end{centering}

\caption{\label{SpherePairProtection}Protective region $\mathcal{P}_{D}$
for the difference walker in the case of two hard disks in two dimensions.
In this case $\mathcal{P}_{D}$ is a cut disk (blue) and the propagator
$c_{D}(\V{r}_{D};t)$ is analytically complex.}

\end{figure}

Unlike the case of cube-shaped protections considered in Ref. \citet{FPKMC1},
protection region of the difference walker for a pair of spherical
particles cannot be a sphere. Among several choices considered, in
our current implementation we opt to protect the difference walker
with a \emph{cut sphere}, i.e. $\mathcal{P}_{D}=\left\{ \V{r}_{D}\mid\left\Vert \D{\V{r}_{D}}\right\Vert <R_{D}^{\mathcal{P}}\mbox{ and }\left\Vert \V{r}_{D}\right\Vert \geq R_{A}+R_{B}\right\} $,
where $\D{\V{r}_{D}}=\V{r}_{D}-\V{r}_{D}^{0}$. Lacking the spherical
symmetry and having more complex geometry of the first-passage surface
(see Fig. \ref{SpherePairProtection}), finding a usable analytical
solution for the propagators in this protective volume is more involved
than for the cubes. Various approximations for the distribution of
exit times and locations $J_{D}(\tilde{t},\tilde{\V{r}}_{D})$ for
cut spheres have been considered in the context of diffusion Monte
Carlo \citet{FP_Diffusion_Spheres,FP_Diffusion_Tabulation}. However,
the previous work used only time-averaged solutions but not the full
time-dependent Green's function $c_{D}(\V{r}_{D};t)$. In the next
section we describe a practical alternative to the full analytical
solution of the first-passage problem inside the cut sphere.

\subsubsection{\label{SectionHopping}Hopping-Based Propagators}

Rather than trying to solve for the exact propagators for the center
walker in a cut sphere, we resort to generating a random walk through
a sequence of small displacements in randomly chosen directions. In
principle, restricting the hops to sufficiently small displacements,
such discrete walks can approximate the samples from the exact continuum
distributions to any desired accuracy. In practice, we use random
walks only for particle pairs that have been brought close to collisions
through the use of exact continuum single-particle propagators. For
such pairs, the walks are typically short and do not entail high computational
cost. 

The time when the trajectory brings the difference walker to the boundary
$\partial\mathcal{P}_{D}$ of the cut sphere $\mathcal{P}_{D}$ is
taken as an approximation for the first-passage time $\tilde{t}$
and the location where the sampled walk hits the surface $\partial\mathcal{P}_{D}$
is an approximation to $\tilde{\V{r}}_{D}$. At every hop a constant
increment $\D{t}_{h}=\D{r}_{h}^{2}/(2D_{D})$ is added to the running
time and the displacements along each dimension are sampled independently
from a one-dimensional Gaussian probability distribution with the
standard deviation $\D{r}_{h}^{2}$. Alternatively, the magnitudes
of walker displacements can be kept equal to $\D{r}_{h}$ while the
time increments can be sampled from the first-passage distribution
$J_{1}$ for the sphere of radius $\D{r}_{h}$. In the limit of small
displacement length both methods should reproduce the required first-passage
probability distributions in the cut sphere.

As a practical matter, one must ensure that the path makes some progress,
that is at least one hop is taken. Therefore $\D{r}_{h}$ should be
smaller than some fraction (e.g., half or one third) of the inter-particle
gap $\delta$. Furthermore, to ensure reasonable accuracy the displacement
$\D{r}_{h}$ should be smaller than the relevant length-scales of
the cut sphere $\mathcal{P}_{D}$. We set $\D{r}_{h}=\epsilon_{h}\min(\delta,R_{A}+R_{B})$,
where $\delta$ is the initial inter-particle gap and $\epsilon_{h}\ll1$
is a fractional \emph{hop length} parameter. The walk terminates when
a collision or pair dissolution occurs, or when time exceeds a specified
upper bound on $\tilde{t}$, $t_{max}$. Occasionally, the first displacement
has to be truncated in order to ensure that the difference walker
does not leave $\mathcal{P}_{D}$ after only one hop: we simply re-sample
the hop again if the initial displacement is too large. In cases when
the particles reflect on collisions, we simply reject the last hop
leading to collision. In cases when the particles react on collisions,
we take the full length of the last hop but truncate the last time
increment. If instead of collision a pair dissolution event occurs,
we reject the last hop rather than trying to truncate it more accurately
(and expensively). In all cases, exit time $\tilde{t}_{D}=t$ and
location $\tilde{\V{r}}_{D}$ are recorded for possible future use.
If the time $t$ exceeds $t_{max}$, the last time increment is set
to $\D{t}_{h}^{'}=t-t_{max}$ and the length of the last hop is scaled
by factor $\sqrt{\D{t}_{h}^{'}/\D{t}_{h}}$.

If and when the previously scheduled first-passage event actually
occurs, we simply move the difference walker to the pre-sampled exit
location $\V{r}_{D}=\tilde{\V{r}}_{D}$. However, when the scheduled
first-passage event is preempted by some other event, e.g. intrusion
of a third party particle, a no-passage propagation is required instead.
In such instances, rather than storing the entire walk generated during
pre-sampling, we opt to repeat the same trajectory starting from the
same seed for the pseudo-random number generator. For this purpose
we store the random seed for each protected pair when its next first-passage
event is scheduled, and use this seed should it become necessary to
repeat the previously sampled walk.

\section{\label{SectionAlgorithmic}Implementation of the FPKMC Algorithm}

In this section we give details of various components of the FPKMC
algorithm, as we have implemented and tested them. It is important
to note that there are alternative implementation choices, some of
which we point to. First, we briefly discuss computational techniques
for searching for near neighbors. We then explain the types of events
that are scheduled and processed by the FPKMC algorithm, and give
a pseudocode for the core of the algorithm, the main event loop. The
components used in the event loop are then briefly described. Finally,
we discuss optimization of the efficiency of the algorithm and an
important direction for future improvement.

\subsection{\label{SectionNeighborSearch}\label{SectionLLC}Near-Neighbor Search}

Efficient particle-based algorithms use various geometric techniques
to reduce to $O(1)$ the cost of searching for the neighbors of a
given particle. Reference \citet{Event_Driven_HE} provides extensive
details and illustrations of these techniques for hard spheres and
ellipsoids; here we briefly summarize the basic linked list cell (LLC)
method and describe the more involved \emph{near-neighbor list} (NNL)
method in Appendix \ref{SectionNNL}. In a certain sense the details
of these geometric techniques are orthogonal to the FPKMC algorithm,
however, efficient neighbor search plays a very important role in
determining the efficiency of the algorithm. Therefore, it is important
to keep in mind how this search is actually performed, especially
when constructing protective regions.

The most basic technique is the so-called \emph{linked list cell}
(LLC) method. The simulation domain, typically an orthogonal box,
is partitioned into $N_{c}$ cells, typically cubes. Each particle
$i$ stores the cell $c_{i}$ to which its centroid belongs, and each
cell $c$ stores a list $\mathcal{L}_{c}$ of all the particles it
contains. Given a particle and a search range, the lists of potential
neighbors is determined by scanning through the neighboring cells.
Typically, for maximal efficiency the cell should be larger than the
largest search range, that is, larger than the largest protective
region $\mathcal{P}_{i}$. If some particle grows too large, e.g.
due to coalescence reactions, one can enlarge the cells and re-build
the associated linked lists. In Section \ref{SectionOptimization}
we discuss the impact of cell size on efficiency as well as methods
for choosing the optimal cell size in simulations of radiation damage.

In our FPKMC implementation, in addition to the list of particles
$\mathcal{L}_{c}$, each cell $c$ stores a \emph{bitmask} $\mathcal{M}_{c}$
consisting of at least $N_{s}$ bits ($N_{s}$ is the number of species).
Bit $\gamma$ in the bitmask $\mathcal{M}_{c}$ is set if cell $c$
contains a particle of species $\gamma$. The bit is set whenever
a particle of species $\gamma$ is added to the cell, and all bitmasks
are cleared periodically. When performing a neighbor search for particle
$i$, cells not containing particles of species that interact with
species $s_{i}$ are easily found and are simply skipped. This can
significantly speed up the neighbor searches in cases where not all
particles interact with all other particles, for example in simulations
of two species diffusion-limited annihilation $A+B\rightarrow0$ where
particles of like species do not interact.

\subsection{Types of Events}

The following types of events, as identified by the event partner
$p$ and event type $\nu$ (which we arbitrarily represent with an
integer here), are scheduled and processed in the FPKMC algorithm
for a given particle $i$ of species $\alpha$:

\begin{description}
\item [{Particle~protection}] identified by $p=0$, $\nu=1$. Such events
are scheduled at the beginning for all particles and immediate updates
are scheduled for any particles whose protections are destroyed before
the scheduled first-passage time. Processing consists of protecting
the particle with a new protective region $\mathcal{P}_{i}$ (see
Section \ref{SectionRecursiveProtection}).
\item [{Particle~update}] identified by $p=0$, $\nu=0$. A protected
particle needs a new event prediction.
\item [{Particle~insertion}] identified by $p=0$, $\nu=-1$. Processing
requires checking whether the newly inserted particle overlaps with
any existing particles. If it does, the corresponding reactions are
processed, otherwise, the particle is protected.
\item [{First-passage~hop}] identified by $p=i$, $\nu<0$. Scheduling
consists of sampling an exit time $\tilde{t}$ and an exit location
$\tilde{\V{r}}$ from $J_{1}$. Processing consists of moving the
particle, $t_{i}\leftarrow t_{i}+\tilde{t}$, $\V{r}_{i}\leftarrow\V{r}_{i}+\tilde{\V{r}}$,
destroying the old protection, and then protecting the particle again.
Additional information, such as the surface of $\mathcal{P}_{i}$
with which the particle collides, can be recorded in $\nu$.
\item [{Particle~decay}] identified by $p=i$, $\nu>0$. The particle
decays via the reaction $\mathcal{D}_{\nu}^{\alpha}$ before leaving
its protection. Scheduling consists of sampling an exponentially-distributed
random number with mean $(\sum_{_{k}}\Gamma_{k}^{\alpha})^{-1}$ and
then choosing one of the reactions $\nu$ with probability $\Gamma_{\nu}^{\alpha}/\sum_{_{k}}\Gamma_{k}^{\alpha}$
(here the sum is over all single particle decay reactions available
for particle \emph{$i$}). Processing consists of sampling from $c_{1}(\V{r};t)$
and moving the particle, $t_{i}\leftarrow t_{i}+t$, $\V{r}_{i}\leftarrow\V{r}_{i}+\V{r}$,
destroying the protection $\mathcal{P}_{i}$, and then executing the
selected decay reaction $\mathcal{D}_{\nu}^{\alpha}$.
\item [{Hard-wall~update}] identified by $p=-w$, $\nu=0$, where $w$
is the identifier of the hard wall. A particle-wall pair needs a new
event prediction after it has been protected.
\item [{Hard-wall~escape}] identified by $p=-w$, $\nu<-1$. This event
is similar to a first-passage hop, however, $J_{HW}$ is sampled instead
of $J_{1}$. Additional information such as the surface of $\mathcal{P}_{i}$
with which the particle collides could be recorded in $\nu$.
\item [{Hard-wall~collision}] identified by $p=-w$, $\nu=-1$. This event
is similar to the hard-wall escape, however, the outcome is a collision
of the particle with the hard wall resulting in annihilation of the
particle (a reflection is never explicitly scheduled).
\item [{Hard-wall~decay}] identified by $p=-w$, $\nu>0$. The particle
decays via the reaction $\mathcal{D}_{\nu}^{\alpha}$ before colliding
with its protection or the hard wall.
\item [{Pair~update}] identified by $p=j$, $\nu=0$. A pair needs a new
event prediction after it has been protected. 
\item [{Pair~disassociation}] identified by $p=j$, $\nu<-1$, where $j$
is the partner particle. Scheduling consists of sampling from $J_{2}(\tilde{t},\tilde{\V{r}}_{i},\tilde{\V{r}}_{j})$.
Processing consists of moving both particles $t_{i}\leftarrow t_{i}+\tilde{t}$,
$\V{r}_{i}\leftarrow\V{r}_{i}+\tilde{\V{r}}_{i}$, and $t_{j}\leftarrow t_{j}+\tilde{t}$,
$\V{r}_{j}\leftarrow\V{r}_{j}+\tilde{\V{r}}_{j}$, destroying both
protections $\mathcal{P}_{i}$ and $\mathcal{P}_{j}$, scheduling
an immediate protection event for particle $j$, and then protecting
particle $i$ again. The specific meaning of the disassociation event
could be recorded in $\nu$, e.g. that particle $j$ left its protection
or that the center walker left its  protection.
\item [{Pair~collision}] identified by $p=j$, $\nu=-1$. Scheduling consists
of sampling from $J_{2}(\tilde{t},\tilde{\V{r}}_{i},\tilde{\V{r}}_{j})$.
Processing consists of executing reaction $\mathcal{R}_{\alpha\beta}$,
where $\beta$ is the species of partner particle $j$.
\item [{Pair~decay}] identified by $p=j$, $\nu>0$. Particle $i$ decays
before the pair disassociates or collides. Processing consists of
sampling $c_{2}(\V{r}_{i},\V{r}_{j};t)$ and moving both particles
accordingly, destroying both protections, scheduling an immediate
protection for particle $j$, and then processing the decay reaction
$\mathcal{D}_{\nu}^{\alpha}$ for particle $i$.
\end{description}

\subsection{Main Event Loop}

The core of the FPKMC algorithm is the event loop described in Algorithm
\ref{FPKMCLoop}. We have already discussed the general framework
and give here additional technical details to facilitate other implementations
of the FPKMC algorithm. For pairs $ij$, we only insert one of the
particles into the event queue (heap), specifically, $\min(i,j)$
if both are mobile or just the mobile particle otherwise. Note that
whenever the position of a particle is updated, its time is updated
as well, $t_{i}\leftarrow t$, and also the LLCs and/or NNLs need
to updated accordingly.

\begin{Algorithm}{\label{FPKMCLoop}FPKMC event loop. Here $\xi$
denotes a uniform random number $0<\xi<1$. Initially the time of
the next particle insertion $t_{\mathcal{B}}=-1$, and all particles
are put in the event queue with $t_{e}=0$, $p=0$, $\nu=1$.}

\begin{enumerate}
\item \label{PopTopOfHeap}Find (query) the top of the event heap to find
the next particle $i$ to have an event with $p$ at $t_{e}$. 
\item If $t_{\mathcal{B}}<0$, set $t_{\mathcal{B}}=-\ln(\xi)/\Gamma_{\mathcal{B}}$,
where $\Gamma_{\mathcal{B}}=\sum_{\alpha=1}^{N_{s}}\mathcal{B}_{\alpha}$
is the total insertion rate of all particle types.
\item If $t_{\mathcal{B}}<t_{e}$, set $t\leftarrow t_{\mathcal{B}}$, $t_{\mathcal{B}}\leftarrow-1$,
choose which particle species $\alpha$ is to be inserted with probability
$\mathcal{B}_{\alpha}/\Gamma_{\mathcal{B}}$ (using linear or binary
(tree) search), insert the new particle into the system, and cycle
back to step \ref{PopTopOfHeap}.
\item Remove particle $i$ from the event queue, store the time increment
$\Delta t=t_{i}^{e}-t_{i}$, the event partner $p=p_{i}$, event type
$\nu=\nu_{i}$, and particle species $\alpha=s_{i}$. Advance the
simulation time $t\leftarrow t_{i}^{e}$.
\item \emph{Insertion}: If $p=0$, $\nu=-1$, check if the newly inserted
particle overlaps with any existing particle and if so, process the
corresponding reactions. If the particle still exists after the check,
set its event to be a regular protection, $p\leftarrow0$, $\nu\leftarrow1$.
Otherwise, cycle back to step \ref{PopTopOfHeap}.
\item \emph{Single-particle event: }If $p=i$ and $\nu\neq0$, update particle
$i$:

\begin{enumerate}
\item \emph{First passage hop}: If $\nu<0$, propagate $i$ to its first-passage
location, $\V{r}_{i}\leftarrow\V{r}_{i}+\tilde{\V{r}}$, and destroy
its protection $\mathcal{P}_{i}$. For spherical particles, one can
sample $\tilde{\V{r}}$ at this point instead of pre-computing it
when scheduling first-passage hops (see Section \ref{SectionMonomers}).
\item \emph{Decay}: Else if $\nu>0$, sample $\V{r}$ from $c(\V{r};\Delta t)$,
set $\V{r}_{i}\leftarrow\V{r}_{i}+\V{r}$, destroy $\mathcal{P}_{i}$,
process the decay reaction $\mathcal{D}_{\nu}^{\alpha}$ and cycle
back to step \ref{PopTopOfHeap}.
\end{enumerate}
\item \emph{Hard-wall event: }Else if $p=-w<0$ and $\nu\neq0$, update
the particle-wall pair $p-w$:

\begin{enumerate}
\item \emph{First-passage event}: If $\nu<0$, propagate particle $i$ to
its first-passage time, and update $\V{r}_{i}$ and $\nu_{i}$ if
needed. \emph{Collision}: If $\nu\leftarrow\nu_{i}=-1$, delete particle
$i$ and cycle back to step \ref{PopTopOfHeap}.
\item \emph{Decay}: Else if $\nu>0$, sample $\V{r}$ from $c_{HW}(\V{r};\Delta t)$,
set $\V{r}_{i}\leftarrow\V{r}_{i}+\V{r}$, destroy $\mathcal{P}_{i}$,
process the decay reaction $\mathcal{D}_{\nu}^{\alpha}$ and cycle
back to step \ref{PopTopOfHeap}.
\end{enumerate}
\item \emph{Pair event}: Else if $p>0$ and $p\neq i$ and $\nu\neq0$,
update the particle pair $ij$, $j=p$, $\beta=s_{j}$. Test whether
the partner is a mobile particle, i.e., whether $p_{j}=i$.

\begin{enumerate}
\item \emph{Disassociation: }If $\nu<0$, propagate the pair to its first
passage time (see Section \ref{SectionPairs}), update $\V{r}_{i}$
and, if needed, $\V{r}_{j}$ and $\nu_{i}$. Destroy protections $\mathcal{P}_{i}$
and $\mathcal{P}_{j}$.
\item \emph{Collision}: If $\nu\leftarrow\nu_{i}=-1$, process the particle-particle
collision. If $j$ is immobile, find and un-protect all of its other
partners $k$ (see Section \ref{SectionMiscSteps}). Process reaction
$\mathcal{R}_{\alpha\beta}$, scheduling immediate particle insertion
events for any remaining or newly created particles, and cycle back
to step \ref{PopTopOfHeap}.
\item \emph{Disassociation}: Else if $\nu<0$, schedule an immediate protection
for the partner, $p_{j}\leftarrow0$, $\nu_{j}\leftarrow1$, and insert
$j$ into the event queue.
\item \emph{Decay}: Else if $\nu>0$, sample $c_{2}(\V{r}_{A},\V{r}_{B};\D{t})$,
set $\V{r}_{i}\leftarrow\V{r}_{i}+\V{r}_{A}$ and destroy $\mathcal{P}_{i}$.
If $j$ is mobile, set $\V{r}_{j}\leftarrow\V{r}_{j}+\V{r}_{B}$,
destroy $\mathcal{P}_{j}$, schedule an immediate protection for the
partner, $p_{j}\leftarrow0$, $\nu_{j}\leftarrow1$, and insert $j$
into the event queue. Process the decay reaction $\mathcal{D}_{\nu}^{\alpha}$
and cycle back to step \ref{PopTopOfHeap}.
\end{enumerate}
\item If particle $i$ is mobile, build a new protection $\mathcal{P}_{i}$
and find the neighbor $k$ that is limiting the size of the protection. 
\item \label{Enlarge_P_i}If $\nu\neq0$ and $k$ is a protected particle,
and $\mathcal{P}_{i}$ is too small (see the discussion in Section
\ref{SectionSummaryOfFPKMC}), then try to enlarge $\mathcal{P}_{i}$
by making more room for it:

\begin{enumerate}
\item Destroy $\mathcal{P}_{i}$ and un-protect $k$.
\item Build a new protection $\mathcal{P}_{i}$ and find the new limiting
neighbor $k'$.
\item Set $k\leftarrow k'$ and cycle back to step \ref{Enlarge_P_i}.
\end{enumerate}
\item Let $p=p_{i}$ (this may have changed during the previous step). If
$p=i$ or $p=0$ then schedule a new single-particle event for $i$:

\begin{enumerate}
\item If $N_{d}^{\alpha}>0$ then sample the time of next decay $t_{d}=-\ln(r)/\Gamma_{d}^{\alpha}$,
where $\Gamma_{d}^{\alpha}=\sum_{_{k}}\Gamma_{k}^{\alpha}$, and find
reaction $\nu$ for which $\sum_{_{k=1}}^{\nu-1}\Gamma_{k}^{\alpha}<r\Gamma_{d}^{\alpha}<\sum_{_{k=1}}^{\nu}\Gamma_{k}^{\alpha}$
using linear or binary search. Otherwise let $t_{d}=\infty$.
\item If $i$ is protected, sample an exit time $\tilde{t}$ and (optionally)
an exit location $\tilde{\V{r}}$ from $J_{1}$ using $t_{d}$ as
an upper bound. Otherwise set $\tilde{t}=\infty$.
\item Choose the smaller of $t_{d}$ and $\tilde{t}$ and insert $i$ in
the event queue with the appropriate event prediction.
\end{enumerate}
\item \label{SchedulePairEvent}Else if $p>0$ and $p\neq i$, schedule
a new pair event for $ij$, $j=p$, $\beta=s_{j}$. If $j$ is mobile,
i.e., if $p_{j}=i$, then set $k=\min(i,j)$, otherwise set $k=i$.

\begin{enumerate}
\item Sample a new decay time for particle $i$, $t_{i}^{d}$ and sample
$t_{j}^{d}$ if $j$ is mobile. Set $t_{k}^{e}$ to the time of the
first decay reaction, set $\nu_{k}$ to the selected decay reaction
and set $k_{d}=i$ or $k_{d}=j$ to indicate the decaying particle.
\item Sample an exit time and location for the pair from $J_{2}(\tilde{t},\tilde{\V{r}}_{i},\tilde{\V{r}}_{j})$
using $t_{k}^{e}$ as an upper bound. If $\tilde{t}<t_{k}^{e}$, change
$t_{k}^{e}\leftarrow\tilde{t}$ and change $\nu_{k}$ appropriately.
If needed, store the exit location.
\item If $\nu_{k}>0$, insert $k_{d}$ into the event queue with the appropriate
decay event prediction. If $k_{d}\neq k$, delete $k$ from the event
queue.
\end{enumerate}
\item Else if $p=-w<0$ then schedule a new event for the particle-wall
pair similarly to the case of an immobile partner in step \ref{SchedulePairEvent}.
\item Cycle back to step \ref{PopTopOfHeap}.
\end{enumerate}
\end{Algorithm}

\subsection{\label{SectionMiscSteps}Steps in the Algorithm}

There are several recurring operations in the FPKMC algorithm that
are implemented as separate subroutines in our Fortran 95 code for
which we do not give detailed pseudocodes but list them below with
brief notes:

\begin{itemize}
\item \emph{Inserting a new particle} of a given species $\alpha$\emph{.}
First the position of the particle is sampled from a problem-dependent
distribution, and the new particle is inserted into the corresponding
cells and/or neighbor lists. The particle is marked as unprotected
and an immediate insertion event ($t_{e}\leftarrow t$, $p=0$, $\nu=-1$)
is scheduled for it.
\item \emph{Protecting a given unprotected particle $i$.} This step is
one of the most complex but also critical to the performance of the
algorithm and is discussed in more detail in Section \ref{SectionRecursiveProtection}.
\item \emph{Un-protecting particle} $i$. The particle $i$ and, potentially,
its mobile partner $j$ are brought to the current time using one
of the propagators $c_{1}$, $c_{2}$ or $c_{HW}$, depending on partner
type $p_{i}$. The LLCs are updated accordingly, $\mathcal{P}_{i}$
and potentially $\mathcal{P}_{j}$ are destroyed, and immediate protection
events scheduled for particle $i$ and potentially its partner $j$.
\item \emph{Un-protecting all partners of an immobile particle}. This is
necessary when the state of an immobile particle changes (e.g., it
decays). Immobile particles may have multiple partners, which are
not stored explicitly, therefore, the partners of the immobile particle
are first identified by performing a neighbor search for any protective
regions $\mathcal{P}_{j}$ that might overlap with $\mathcal{C}_{i}$
and checking if $p_{j}=i$. These partners are then unprotected and
immediately scheduled for re-protection. Note that the neighbor search
here relies on the LLCs and it is not necessarily safe to modify LLCs
until the search completes.
\item \emph{Scheduling and processing of pair events}. The implementation
depends on the particle shapes and the types of reactions considered.
Section \ref{SectionPairPropagators} discusses pair propagators for
hard spheres.
\item \emph{Processing a collision between two particles}. This step is
very application-specific because of the different types of reactions
that may occur. Typically the processing involves deleting some particles
and then possibly inserting others.
\item \emph{Processing a decay reaction}. This is also application-specific
and consists of deleting the decaying particle and then inserting
the reaction products at desired positions.
\item \emph{Resetting the time counter to} $t=0$. This step is useful for
minimizing round-off errors, especially before an event generating
other events with very small timestamps occurs. For example, insertions
of cascades of defects creates dense lumps of particles that evolve
at time scales $\D{t}$ comparable to numerical precision ($10^{-16}$)
relative to the time scales of the majority of events, and thus $t+\D{t}\approx t$
due to round-off. This can be avoided by setting $t=0$ after subtracting
the current $t$ from all time counters, including the particle times
$t_{i}$ and the event predictions $t_{i}^{e}$.
\item \emph{Synchronizing all particles}. It is occasionally useful to bring
the whole system to the current point in time for analysis, saving
the configuration to a file, etc. This requires un-protecting all
particles. This is a good occasion to reset the current time $t\leftarrow0$
to avoid the round-off problems.
\end{itemize}

\subsection{\label{SectionRecursiveProtection}Particle Protection}

For better performance, one can try to use the freedom afforded by
the FPKMC algorithm to select particle protection so as to delay as
far as possible the very next event in the queue. However, finding
an optimal space partitioning is a difficult problem of non-linear
optimization, especially since events other than the first-passage
and no-passage propagations are taking place concurrently during the
simulation. As described in Section \ref{SectionSummaryOfFPKMC},
Algorithm \ref{FPKMCLoop} takes the strategy of first trying to protect
a given \emph{seed} particle $i$ with the largest possible protection
without disturbing other, protected, particles. Then, some of those
other particles may be unprotected to make room and the process is
repeated. In this section we describe our procedure for protecting
an unprotected particle. Note that when this procedure is invoked,
there can be an arbitrary number of other unprotected particles.

Our algorithm for protecting a given particle $i$ finds the nearest
pair of unprotected particles whose protection affects the protection
of $i$. It is recursive and rather complex in its details; here we
try to give a more intuitive and brief verbal explanation that can
be used to design alternative implementations. The algorithm starts
from the particle $i$ and finds the maximal possible size of its
protective region by examining all of the neighboring objects limiting
the protection and finding the {}``nearest'' (most-limiting) neighbor
and also the {}``next-nearest'' (next-limiting) neighbor. These
neighbors could be other protected and unprotected particles, nearby
hard walls, the cells used to build the LLCs/NNLs, etc. For each of
these cases one can calculate the maximal allowed size of the protection
$\mathcal{P}_{i}$ afforded by the neighboring object. Specifically,
if the neighboring object is itself an unprotected particle it is
assumed that the particles would be protected with touching protections
whose sizes are proportional to the square roots of the diffusion
coefficients. If the limiting neighbor is an unprotected particle,
the algorithm recurses by repeating the process with that particle
replacing particle $i$. The recursion continues until a neighbor
is found whose own limiting neighbor is the particle $i$, that is,
a pair of mutual nearest neighbors $i$ and $j$ is found. If $i$
and $j$ are sufficiently close and the next-nearest neighbors of
$i$ and $j$ allow for pair protection (see Section \ref{SectionPairs}),
the particles $i$ and $j$ are protected as a pair. Otherwise, they
are protected with touching single-particle protections. The recursion
trail is then reversed and all the particles visited during the forward
recursion are protected with the maximal allowed protection, accounting
for the protection of their nearest neighbor during the forward pass
of the recursion and also reusing the previously-identified next-nearest
neighbor. At the end of the process, the particle $i$ and possibly
a number of other particles are protected and the neighbor limiting
the size of $\mathcal{P}_{i}$ has been identified.

\subsection{\label{SectionOptimization}Optimizing Runtime Parameters for Efficiency}

The most important parameter defining the performance of the FPKMC
algorithm is the size of the cells used for neighbor searches. The
common wisdom for a homogeneous system of identical particles is that,
optimally, there should be about one particle per cell which balances
the cost of neighbor searches with the cost of updating the LLCs and
moving particles between the cells. However, this prescription does
not necessarily apply to the often heterogeneous (in both time and
space) systems encountered, for example, in radiation damage simulations.
Some clusters of defects can grow to sizes more than then times larger
than monomer defects. Additionally, the system's evolution can entail
disparate timescales differing by many orders of magnitude, from fast
relaxation on the scale of picoseconds during initial cascade insertion
to slow annealing on the scale of years.

For simplicity, the following discussion focuses on LLCs, without
NNLs. Furthermore, we assume that among several particle species present
in the system there is a single highly mobile species $\alpha_{m}$,
such as interstitials in radiation damage modeling. In such cases,
the bulk of computational effort is spent on protecting and propagating
particles of the fast species. Generally, it is a good idea to assign
as large protection as possible for the fast particle(s) without,
however, having to search for too many neighbors while building a
protective region. In our algorithm the protection size is limited
by the smaller of the following: 

\begin{enumerate}
\item The range for the neighbor search $R_{s}=\min(n_{s}L_{c}/2,$$n_{s}L_{c}-R_{max})$,
where $n_{s}=\left\lceil (R_{max}+R_{p})/L_{c}\right\rceil $ is the
number of neighboring cells to be searched in each direction, $L_{c}$
is the linear dimension of the (cubic) cells, $R_{p}$ is the radius
of particle species $\alpha_{m}$, and $R_{max}\gg R_{p}$ is the
current maximum size of protection in the entire system (typically
an immobile cluster).
\item The distance $R_{v}$ to nearby neighbors against which pair protection
is not possible (due to third particles blocking it or the neighbor
being too far away to make pair protection advantageous). This is
a measure of the void size around the fastest particles.
\end{enumerate}
The optimal performance is achieved when the two bounds are approximately
equal, $R_{s}\approx R_{v}$, i.e., the number of searched neighbors
is just enough to find the largest possible protection, no more, no
less. The void size $R_{v}$ can be measured for a given configuration
by gradually enlarging the cells until the average size of protective
domains stops increasing, becoming close to $R_{v}$. Numerical tests
have confirmed that indeed the choice of $L_{c}$ such that $n_{s}=1$
and $R_{s}\approx R_{v}$ is optimal.

In an actual simulation it is too expensive to estimate $R_{v}$ at
every step but it is still possible to use an adaptive method for
selecting a cell size. Specifically, we monitor how many protections
for particles of species $\alpha_{m}$ have been limited by cell size
(i.e., by $R_{s}$) and how many have been limited by nearby particles
(i.e., $R_{v}$). We observe that it is best to keep the former a
small but nonzero fraction of the latter. If the runtime statistics
show that too many protections are blocked by the cell size, the cells
are enlarged by reducing the number of cells by one along each dimension
of the simulation volume. Conversely, the cell count is increased
by one if the statistics show that too few protections are blocked
by the cells. The cell size strongly affects the performance of the
code. If too small, the protections will be small too leading to shorter
scheduled propagation times and, thus, slower time evolution. If the
cells grow too large, there will be many neighbors to examine during
each protection, slowing down the calculations. The use of NNLs and,
in particular, BSCs (see Section \ref{SectionNNL}), becomes advantageous
when very large clusters are present, so that $n_{s}=1$, i.e., $L_{c}>R_{max}+R_{p}$
and the cells contain many smaller particles at sufficiently high
monomer densities. In the simulations of irradiated materials reported
in section \ref{SectionRadiationDamage}, we focus on low particle
densities and find that the use of NNLs is not necessary to achieve
an optimal performance.

Note that computational cost is not always dominated by a single ultra-fast
species. As an example, consider the case of modeling radiation damage
inflicted in the form of defect pairs consisting of a very mobile
interstitial and a much less mobile vacancy. At first, the interstitial
propagation events will dominate the event loop and the interstitials
will quickly diffuse to absorbing sinks, such as hard-wall boundaries
or nearby vacancies, and disappear. This will leave behind the slower
vacancies that will continue their random motion until the next defect
pair is inserted. If the insertion rate is low, the vacancies can
move significantly between successive insertions and after the fast
interstitials all died out. During such intervals the computational
cost is dominated by vacancy propagations and protections. In such
conditions, the focus of particle protection will have to shift from
interstitials to vacancies and back to interstitials. In general,
the choice of optimal cells and protection sizes is complex and problem-dependent.
Our implementation of the FPKMC algorithm collects statistics that
can be used to make runtime adjustments and improve the simulation
performance.

\subsection{\label{SectionTimeDriven}Mixing Time-Driven with Event-Driven propagations}

Under certain conditions the exact event-driven handling of particle
diffusion may become inefficient and/or cumbersome. For example, in
a dense group of closely spaced particles, protection of single particles
and pairs is severely limited by the third particles in close proximity.
In such conditions, particle displacements can become too small to
deserve asynchronous event-driven handling. Time-stepping avoids the
cost of event queue operations and simplifies overlap detection. Therefore,
it can be more efficient to use time-stepping for dense groups of
particles, similar to the time-driven hopping algorithm presented
in Section \ref{SectionHopping} that avoids the use of complex pair
propagators for spheres. Use of small hops is also advantageous when
particles or surfaces (e.g., grain boundaries) have complex shapes
making analytical treatment of particle diffusion and collisions difficult
or impossible. Yet another example when simple time stepping is useful
is tightly-bound collections of particles (clusters) that may act
as a single particle and have complex internal structure and dynamics
(relaxation). Adding a time-driven component to the asynchronous event-driven
FPKMC algorithm allows to retain the overall algorithm efficiency
even in such difficult conditions. 

The particles which cannot be protected by a sufficiently large protection
are marked as \emph{time-driven particles} and are \emph{not} inserted
into the event queue. A special type of event, a \emph{time step event},
is introduced and always scheduled to occur at equal time intervals
$\D{t}$. When a time-step event is processed, all time-driven particles
are moved simultaneously, followed by overlap checks and reactions,
if any are detected. Particles that are time-driven are not protected
against each other, instead, they can stay unprotected or are protected
only against the event-driven particles. Time-driven particles whose
protections overlap form a cluster and are propagated synchronously
with the same time step $\D{t}$ tailored to the fastest particle
in the cluster. For simplicity, all time-driven particles may be treated
as one cluster with a single global $\D{t}$. However, it is often
the case that different species have widely differing diffusion coefficients
and therefore very different time-steps will be appropriate for different
species. To solve this problem, one can use the $n$-fold (BKL) \citet{Ising_BKL}
synchronous event-driven algorithm inside each cluster and replace
the time step events with \emph{BKL hop events}. At each hop event
all particles of a single species move by a small but non-negligible
distance while all other particles remain in place. This way, the
more mobile particles move more frequently (with correct relative
frequencies) than the less mobile ones.

Note that the particles in such a time-driven cluster can take hops
up to the time of the next event in the queue, since it is known that
the hops can not be preempted by another event. In some situations
this may improve efficiency by reducing the number of heap operations
and also increasing the memory access locality of the code by focusing
multiple events on the same (cached) small group of particles. We
are currently developing an implementation of a mixed event-driven
(first-passage) with time-driven ($n$-fold KMC) algorithm and will
report additional details and results in a future publication.

\section{\label{SectionResults}Validation and Results}

In this section we apply the FPKMC algorithm to several diffusion-reaction
problems of increasing complexity. To validate the new algorithm and
to demonstrate its efficiency, we compare our simulations to results
obtained using two different object KMC (OKMC) codes developed earlier
for simulations of continuum diffusion (BIGMAC code \citet{BIGMAC})
and for simulations of random walks on a lattice (LAKIMOCA code \citet{LAKIMOCA_OKMC}).
Presented in Section \ref{Section_AB}, the first two test problems
are relatively simple validation studies for the case of two-species
annihilation, $A+B\rightarrow0$, when the two species have different
diffusion coefficients. FPKMC simulations for this model are compared
against results obtained with the BIGMAC code. In Section \ref{SectionOKMC},
we apply FPKMC to a more challenging test problem of damage accumulation
in a metal thin film subjected to electron irradiation. We base our
FPKMC simulations on a well known model of $\alpha$-iron studied
earlier using three different methods, including cluster dynamics,
lattice OKMC, and the approximate continuum OKMC algorithm JERK \citet{JERK_CD_Barbu}.
Here we compare our results for this model against simulations performed
using the LAKIMOCA lattice-based code. Finally, in Section \ref{SectionReactor}
we apply FPKMC to simulations of radiation damage accumulation over
previously inaccessible time scales, namely, to time intervals and
radiation doses characteristic of material lifetimes in a nuclear
reactor. To our knowledge, this is the first time an atomistic model
has reached technologically relevant radiation doses exceeding the
previous simulation benchmarks by several orders of magnitude.

\subsection{\label{Section_AB}Two-Species Annihilation}

As a validation study, let us first consider a system of spherical
particles of two species $A$ or $B$ in three dimensions. The particles
of different species have different radii $R_{A}$ and $R_{B}$ and
different diffusion coefficients $D_{A}$ and $D_{B}$, $D_{B}<D_{A}$.
Particles of like species do not see each other but particles of unlike
species annihilate upon hard-sphere contact, i.e. at the annihilation
distance $r_{A+B\rightarrow0}=R_{A}+R_{B}$. As a base for comparison,
we first simulated the same model reaction using the BIGMAC code for
several values of the hop distance $\Delta$, characterized hereafter
by the dimensionless ratio $\delta=\Delta/(R_{A}+R_{B})$. For the
hopping-based pair propagators in FPKMC we set $\delta=0.1$, which
was found to be sufficiently small to give accurate results, yet large
enough to make the pair propagators almost as efficient as the analytical
pair propagators for cube-shaped particles.

\begin{figure*}
\includegraphics[width=0.49\textwidth]{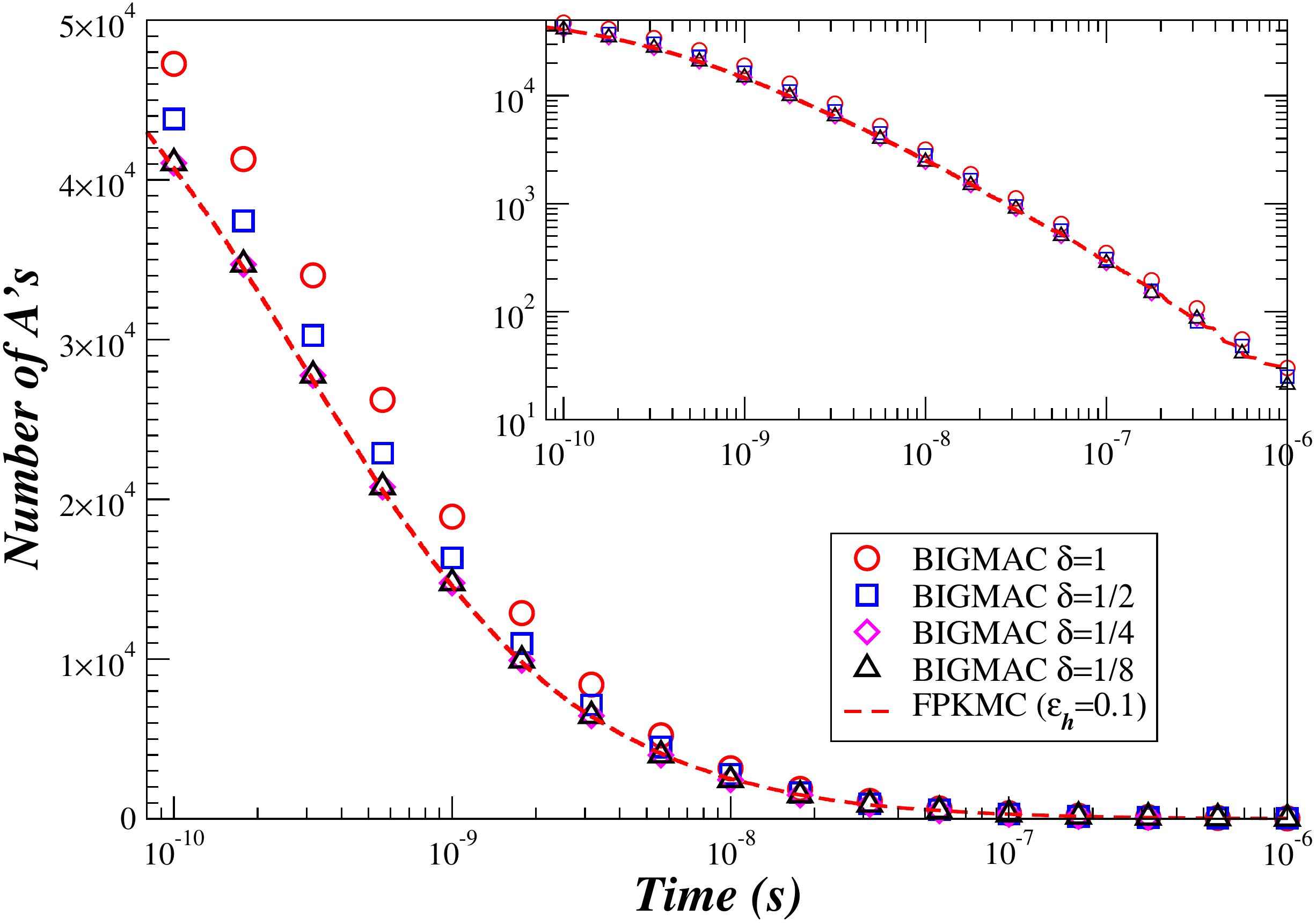}\includegraphics[width=0.49\textwidth]{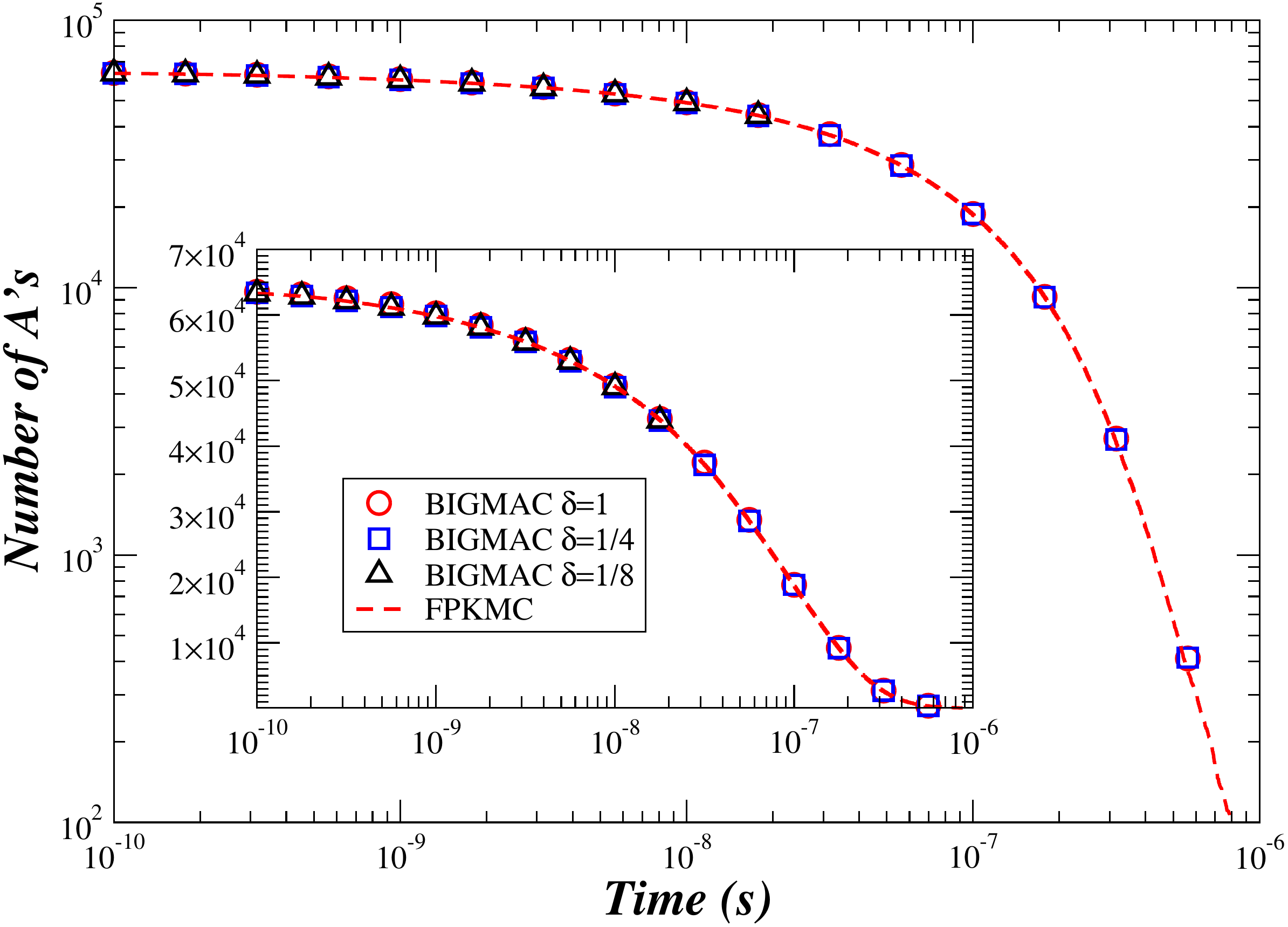}

\includegraphics[width=1\textwidth]{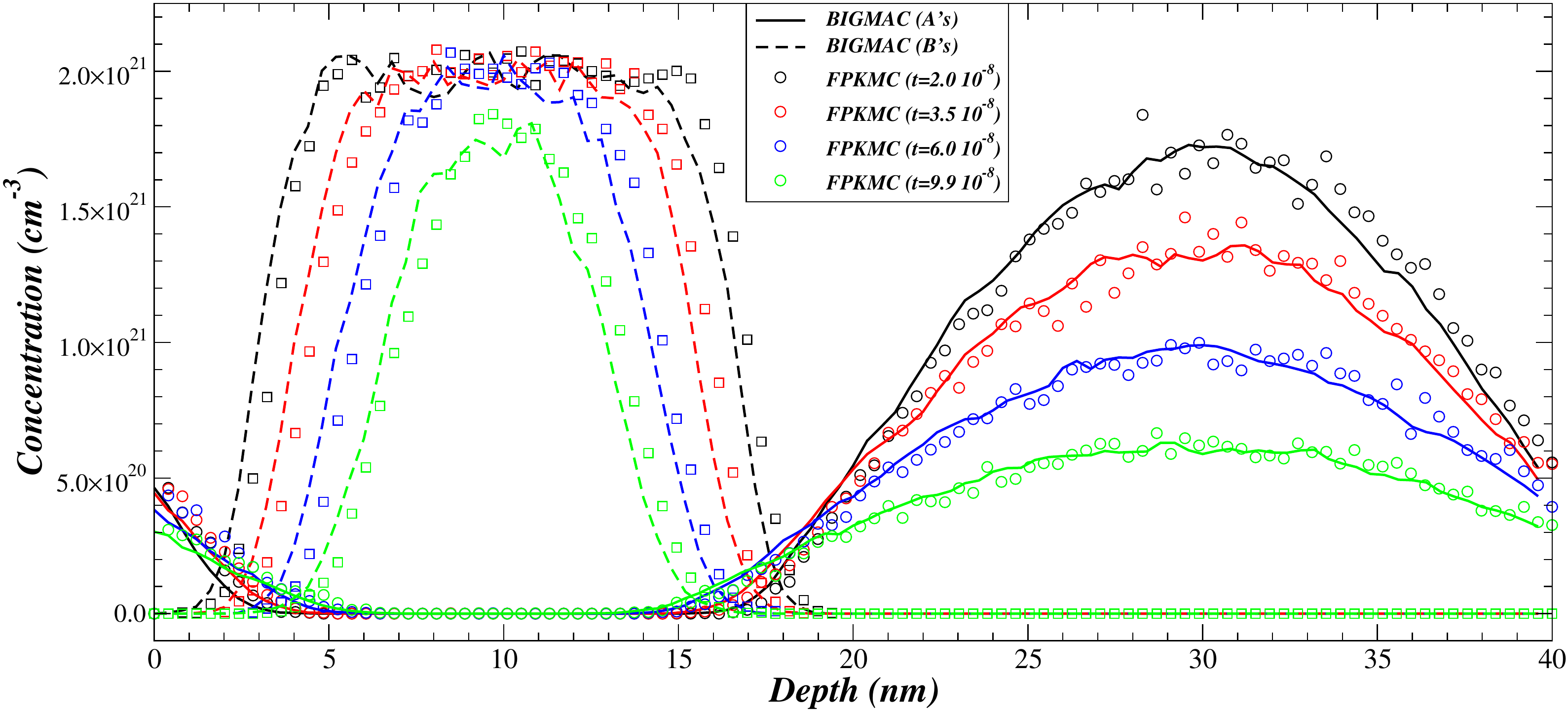}

\caption{\label{A_B_BIGMAC}Comparison between BIGMAC and FPKMC for the two-species
annihilation ($A+B\rightarrow0$) test problem in three dimensions,
starting from a uniformly-mixed (top left) and phase-separated (top
right and bottom) initial state. The two top panels and their insets
show the number of particles as a function on time, $N_{A}(t)=N_{B}(t)$,
using log-log and semi-log scales. The bottom panel shows the particle
concentration profiles $c_{A}(x;t)$ (solid curves for BIGMAC, circles
for FPKMC) and $c_{B}(x;t)$ (dashed curves for BIGMAC, squares for
FPKMC) for the phase-separated initial conditions. Concentration profiles
taken at several times during the simulation are plotted in different
colors. }

\end{figure*}

The simulations were performed in a cubic simulation domain of volume
$L^{3}$ with periodic boundary conditions. In the simulations reported
here, half of the particles are of species $A$ and the other half
are of species $B$. We consider two different initial conditions.
In the first case $A$s and $B$s are randomly and uniformly distributed
in the simulation volume (the overlapping particles of unlike species
are removed). The reaction kinetics is described by the reduction
of the number of $A$ (or $B$) particles with time, $N_{A}(t)=N_{B}(t)$.
Figure \ref{A_B_BIGMAC} shows this decay kinetics for FPKMC simulations
as well as for BIGMAC runs with different hop sizes $\delta$. The
results indicate that hop sizes as large as $\delta=1/4$ can be used
in BIGMAC without a noticeable error and that BIGMAC and FPKMC produce
virtually indistinguishable results.

In the second test case, one half of the box ($x\geq L/2$) is randomly
filled with $A$'s and the other half ($x<L/2$) with $B$'s. The
resulting $N_{A}(t)=N_{B}(t)$ is shown in Fig. \ref{A_B_BIGMAC},
with similar qualitative behavior as in the first case of intermixed
$A$s and $B$s. The concentration profiles $c_{A}(x;t)$ and $c_{B}(x;t)$
are shown for several different points in time in Fig. \ref{A_B_BIGMAC}
for both FPKMC and BIGMAC with hop size $\delta=1/4$ revealing an
excellent agreement between the two algorithms. The efficiency of
BIGMAC simulations is proportional to the square of the hop size,
but even for the rather large hop sizes used in our BIGMAC simulations,
the latter take several hours to annihilate most of the initial $1.28\cdot10^{5}$
particles, whereas the FPKMC code accomplishes the same in about 10
minutes on a typical single-processor desktop machine essentially
independently of the value of $\delta$.

\subsection{\label{SectionRadiationDamage}Radiation Damage}

Modeling of radiation damage in reactor materials is not only another
test problem for our algorithm but also a technologically important
application, notably in the design and maintenance of nuclear power
plants. When a metal is irradiated, incoming high energy particles
(neutrons, ions, electrons) collide with atoms of the host crystal
lattice inducing displacement cascades and producing numerous defects,
such as excess vacancies and interstitials. Many of these defects
quickly annihilate with each other, but some diffuse away from the
initial impact locations, and eventually finding other defects to
react with and to form defect clusters. The cluster density and sizes
can grow over time resulting in substantial (most often detrimental)
modifications of material microstructure and properties. The atomistic
KMC method is well suited for simulations of radiation damage accumulation
by tracing the numerous diffusive hops and reactions among crystal
defects induced by collision cascades. Unfortunately, the same detailed
nature of KMC simulations makes them computationally demanding and
limits their time horizon to times far shorter than the technologically
relevant time scales (years). 

The FPKMC algorithm seems to be perfect for simulations of materials
under irradiation and, in fact, it was this particular application
that served as an initial inspiration for the new method development.
The advantage of FPKMC over other KMC methods is that the new algorithm
handles equally efficiently both the fast stages of cascade annealing
when the density of diffusing defects is high and the slow evolution
of the small number of surviving defects that continue their diffusive
motion during the intervals between subsequent collision cascades.
This intermittent (fast-slow-fast-...) character of system's evolution
is the major computational challenge in radiation damage simulations
that the asynchronous FPKMC algorithm addresses.

To enable radiation damage simulations, several particle species are
introduced: \emph{monomers}, including highly-mobile interstitials
($I$) and less-mobile vacancies ($V$), a number of mobile \emph{cluster
species}, for example, dimers ($I_{2}$ and $V_{2}$) and trimers
($I_{3}$ and $V_{3}$), immobile species representing clusters larger
than any of the mobile species ($I_{c}$ and $V_{c}$), and defect
(Frenkel) pairs ($IV$). Each particle is assigned a hard-sphere radius:
for the clusters the radius is related to the number of monomers $c\geq1$
contained in the clusters assuming that the monomer volumes are additive,
i.e. $R_{c}\sim R_{0}+(R_{1}-R_{0})c^{1/3}$.

Frenkel pairs ($IV$) are inserted in the simulation volume at a specified
birth rate and instantly (i.e., with decay time $\tau_{IV}=0$) decay
via $IV\rightarrow I+V$. The resulting interstitial and vacancy monomers
are placed randomly within the simulation box, either at some initial
distance from each other or completely independently of each other.
Electron irradiation creates individual Frenkel pairs while irradiation
by high-energy ions or neutrons creates damage in the form of \emph{displacement
cascades} producing compact collections of monomers and clusters,
each cascade containing about 100 Frenkel pairs. The collision cascades
are randomly selected from a library of cascade configurations generated
in Molecular Dynamics simulations and the cascade locations and orientations
are sampled from appropriate distributions. 

Upon collisions, particles of like species coalesce, for example,
$I+I\rightarrow I_{2}$ or $V+V_{3}\rightarrow V_{c=4}$, whereas
collisions of particles of unlike species lead to complete or partial
annihilation, for example, $I_{2}+V_{c=4}\rightarrow V_{2}$. In our
current implementation, the distance at which two particles collide
must be equal to the sum of their radii since the FPKMC algorithm
handles the geometry of the protective regions assuming that an additive
hard-sphere interactions among the particles. For consistency, the
same is assumed in simulations performed with the lattice-based LAKIMOCA
code that are used here for comparison. To imitate the stronger effect
of elastic strain on the interstitials (the bias), the interstitials
are assigned radius $1.2$ times larger than that of the vacancies.

The defect clusters emit monomers at a given rate, represented as
a decay reaction, for example, $V_{c=5}\rightarrow V_{c=4}+V$, or
$I_{2}\rightarrow I+I$. The emitted monomer is placed at a preset
\emph{emission distance} $\delta_{e}$ from the cluster surface, in
a random direction. This is at variance with most other commonly adopted
non-local emission rules, e.g. in the mean-field Cluster Rate Theory
method \citet{ClusterDynamics_Review}, in which the monomers are
emitted from the clusters to {}``infinity''. If and when a cluster
shrinks by emission to a size at which it becomes mobile, the species
of the remaining cluster is changed accordingly, for example, $V_{c=4}\rightarrow V_{3}+V$.

The rates of defect diffusion and monomer emission from the clusters
are calculated according to the standard expressions for the rates
of thermally activated rate processes in solids. Defect migration
and binding energies needed to compute the rates are calculated atomistically,
e.g. from first-principles theory, or estimated from experimental
data. In the simple model we consider here, only the monomers are
mobile, with a diffusion coefficient \[
D_{1}=D_{0}e^{-E_{m}/kT},\]
where $E_{m}$ is the activation energy for defect migration (lattice
hop). The rate of emission of monomers from a cluster composed of
$c$ monomers is

\[
\Gamma_{c}=\Gamma_{0}D_{1}a^{-2}c^{2/3}e^{-E_{b}(c)/kT},\]
where $a$ is the lattice spacing, $\Gamma_{0}$ is a constant that
depends on the lattice type and $E_{b}(c)$ is the monomer binding
energy in a cluster of size $c$, estimated using\[
E_{b}(c)=E_{f}+\left[E_{b}(2)-E_{f}\right]\frac{c^{2/3}-(c-1)^{2/3}}{2^{2/3}-1},\]
where $E_{f}$ is the monomer formation energy \citet{JERK_CD_Barbu}.

\subsubsection{\label{SectionOKMC}Thin film of metal under electron irradiation}

As a test problem, here we consider a model previously studied in
Ref. \citet{JERK_CD_Barbu} using two other KMC algorithms and the
mean-field cluster dynamics \citet{ClusterDynamics_Review}. The model
system is a $0.287\mu m$-thick film of $\alpha$-iron subjected to
electron radiation. Periodic boundary conditions are used in the $x$
and $y$ directions, while absorbing walls are used in the $z$ direction.
Further details of the model are given in Ref. \citet{JERK_CD_Barbu}
and will not be repeated here. We made a few minor changes to the
model parameters presented in Ref. \citet{JERK_CD_Barbu}, notably
switching from the non-local to local emission of monomers from the
clusters. Here we compare our FPKMC simulations with new results obtained
using the LAKIMOCA code \citet{LAKIMOCA_OKMC}.

In pure iron, the interstitials are much more mobile than the vacancies
which results in their rapid absorption at the free surfaces or annihilation
with the vacancies. Due to this very high monomer mobility, very few
if any interstitial clusters form. However, the ones that do nucleate,
can grow rather large because the binding energy of interstitials
in the clusters is rather high. A large fraction of computational
effort is therefore expended on propagating the interstitials. However,
depending on the irradiation rate (fluence), the vacancies may have
time to move significantly between successive insertions of Frenkel
pairs. Here we present a high-fluence case studied computationally
and experimentally in Ref. \citet{JERK_CD_Barbu}. Further computational
tests of the FPKMC code in a wide range of irradiation conditions
will be discussed in Section \ref{SectionPerformance}.

\begin{figure*}
\begin{centering}
\includegraphics[width=0.65\textwidth]{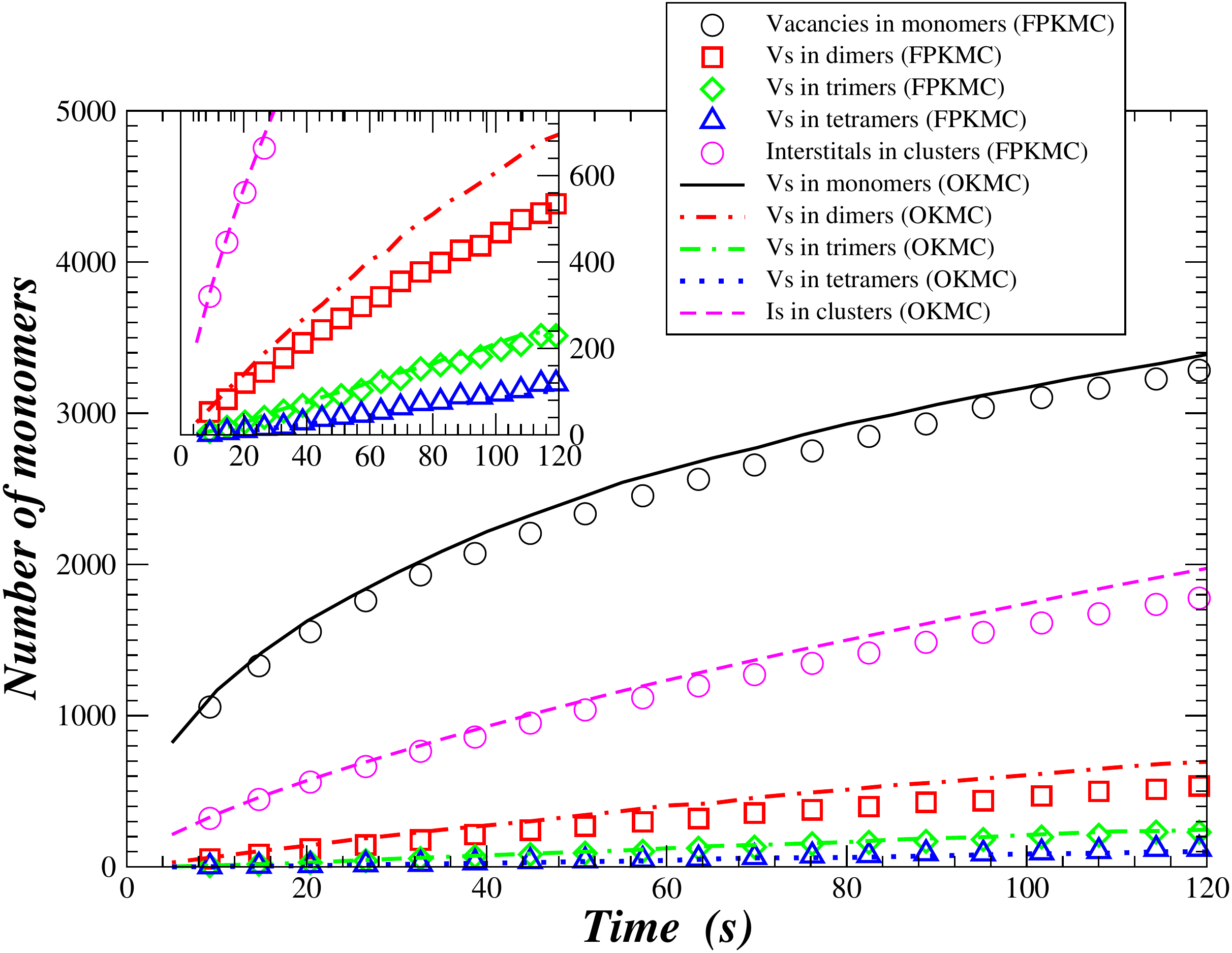}
\par\end{centering}

\begin{centering}
\includegraphics[width=0.65\textwidth]{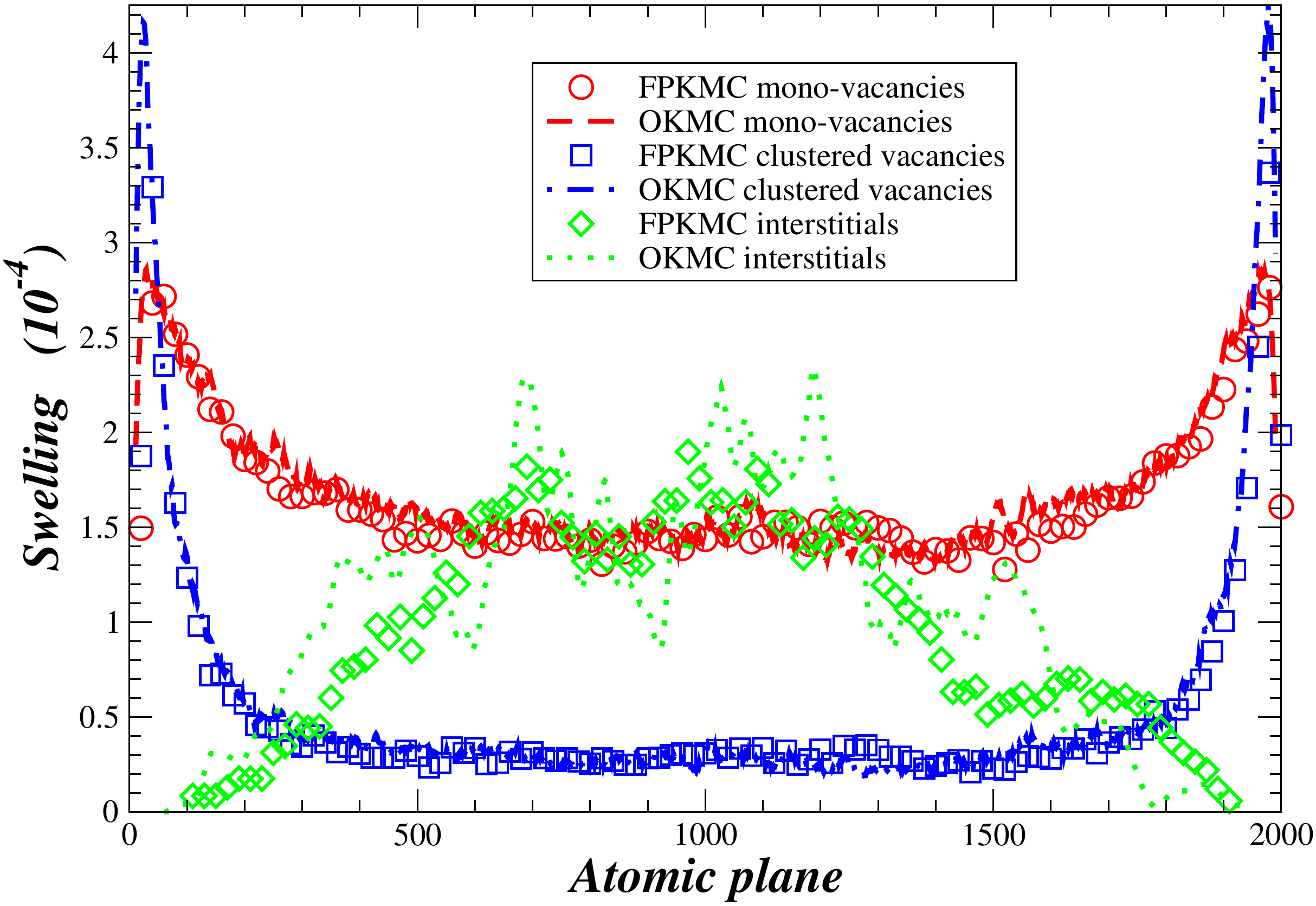}
\par\end{centering}

\caption{\label{BIGMAC_comparison}Comparison between FPKMC (symbols) and LAKIMOCA
\citet{LAKIMOCA_OKMC} (lines) simulations of a $0.287\mu m$-thick
film of $\alpha$-iron subjected to 120 seconds of electron radiation
at a temperature $T=200^{o}C$. The results shown on the plots are
obtained by averaging over 50 runs. (Top) The time evolution of the
total number of mono-vacancies, small vacancy clusters and interstitials
(in clusters). The statistical error bars are comparable to the symbol
sizes. The inset shows a different scale to focus on the smaller clusters.
(Bottom) The density profile along the thickness of the film for all
vacancies (in monomers and clusters), mono-vacancies, vacancies in
clusters and all interstitials (in monomers and clusters) at the end
of 120 seconds of simulated irradiation. The error bars are comparable
to the symbol sizes, except for the interstitials for which the statistics
is poor.}

\end{figure*}

A comparison of the simulated time evolution for the number of defects,
as well as their spatial distribution at the end of 120 seconds of
irradiation, is shown in Fig. \ref{BIGMAC_comparison}. Generally,
we observe a good agreement between FPKMC and LAKIMOCA results although
the number of accumulated defects is slightly smaller in the FPKMC
simulations than in the LAKIMOCA runs. This is perhaps a consequence
of different treatment of defect diffusion in the two methods: in
FPKMC the defects move by continuous diffusion whereas in LAKIMOCA
the defects walk on a lattice. However, at sufficiently low densities,
lattice discreteness is not as important and the effective reaction
and emission rates are well matched between the discrete and continuum
models. Further comparisons between FPKMC and OKMC simulations will
be given in a future publication, here we simply observe the ability
of the FPKMC algorithm to correctly simulate radiation damage by comparing
it to a standard OKMC. Each FPKMC simulation sample used for obtaining
the data plotted in Fig. \ref{BIGMAC_comparison} has taken less than
five minutes to complete on a modest workstation (a $3.4\mbox{ GHz}$
Xeon 64-bit processor). Efficiency of FPKMC simulations is further
discussed in the next section.

\subsubsection{\label{SectionPerformance}Performance}

For a large number of particles $N$, the computational complexity
of the FPKMC algorithm per event should be order $O(\log N)$ which
is the cost of event queue updates. For a typical value of $N=10^{5}$,
the logarithm is masked by other dominant costs that are all linear
(e.g., neighbor searches and sampling from the propagators). Computational
tests in the range of $N$ from $10^{4}$ to $10^{6}$ have indeed
verified that the cost of FPKMC simulations scales linearly with the
number of particles. 

The FPKMC code performance also depends on a number of other parameters,
e.g. particle density, the disparity of diffusion and emission rates,
differences in particles sizes between the different species, etc.
These and other factors and their interactions affect the overall
performance in complex ways that are yet to be fully examined. We
defer to future applications to study the subtle effects of various
model parameters on the method's performance. Here we present a few
figures on the performance of our FPKMC code for the same simple model
of metal thin film described in Section \ref{SectionRadiationDamage},
as a function of irradiation flux measured in the units of displacements
per atom ($dpa$) per second ($dpa/s$). The kinetics of defect microstructure
evolution and, in particular, the amount and character of accumulated
damage depend sensitively on flux and temperature. For example, the
defect density (accumulated damage) increases with the increasing
flux and/or the decreasing temperature. It is therefore not \emph{a
priori} obvious that FPKMC will be equally effective in dealing with
a wide range of fluxes and temperatures. A reasonable measure of the
algorithm performance is the damage dose simulated over a unit of
CPU time expressed for example, in the unit of dose simulated in one
day of computing ($\mbox{dpa}/\mbox{cpuday}$).

\begin{table}
\begin{centering}
\begin{tabular}{|c|c|c|c|c|}
\hline 
Dose rate ($\mbox{dpa/s}$) & Total dose ($\mbox{dpa}$) & Simulated time  & Speed ($\mbox{s/cpus}$) & Efficiency ($\mbox{dpa/cpuday}$)\tabularnewline
\hline
\hline 
$1.5\cdot10^{-4}$ & 18 & 33 hrs & 0.14 & 1.8\tabularnewline
\hline 
$1.5\cdot10^{-5}$ & 2.9 & 54 hrs & 1.3 & 1.7\tabularnewline
\hline 
$1.5\cdot10^{-6}$ & 4.1 & 31 days & 13 & 1.6\tabularnewline
\hline 
$1.5\cdot10^{-7}$ & 1.6 & 125 days & 150 & 2.0\tabularnewline
\hline 
$1.5\cdot10^{-8}$ & 10 & 21 years & $2.1\cdot10^{3}$ & 2.7\tabularnewline
\hline 
$1.5\cdot10^{-9}$ & 8.4 & 175 years & $2.3\cdot10^{4}$ & 3.0\tabularnewline
\hline
\end{tabular}
\par\end{centering}

\caption{\label{FPKMC_performance}Performance of the FPKMC algorithm in simulations
of a $0.287\mu m$-thick film of $\alpha$-iron at $T=473^{o}K$ subjected
to different fluxes (dose rates) of electron radiation. Conditions
typical of nuclear reactors correspond to dose rates on the order
of $10^{-8}\mbox{dpa/s}$ and lifetimes from years to several decades,
while in the accelerated irradiation facilities the dose rates can
be on the order of $10^{-4}\mbox{dpa/s}$ and the tests can last for
several hours or days. A displacement dose of $K-$dpa means that,
on average, each atom has been displaced from its equilibrium lattice
position $K$ times due to incoming radiation. Note that simulation
efficiency changes over the course of one simulation so that the overall
efficiencies reported in the last column are representative averages.}

\end{table}

The figures presented in Table \ref{FPKMC_performance} demonstrate
that, with the use of optimization techniques discussed in Section
\ref{SectionOptimization}, the algorithm performance remains nearly
constant across several decades of radiation flux. This highly desirable
property derives from the ability of our asynchronous event-driven
algorithm to deal with very large differences in event rates and local
densities. We observe that, in the course of a single simulation,
the FPKMC algorithm self-adjusts to the current conditions of spatial
and temporal heterogeneity without much intervention or parameter
tuning.

\subsubsection{\label{SectionReactor}Radiation Damage on Reactor Timescales}

Future development of nuclear energy demands materials that can withstand
harsh conditions of particle irradiation, high temperature, mechanical
stress and active chemical agents over tens of years. The only fully
reliable method to evaluate the potential of a candidate material
is to subject it to conditions relevant for the future reactor designs.
However, such an approach is not practical given that the relevant
environment can be achieved only after the reactor is already built.
Furthermore, even if an appropriate material testing facility were
to exist, testing the candidate materials over the intervals of 50
or 100 years would not be practical.

The idea of \emph{accelerated material testing} is to subject candidate
materials to conditions even harsher than in the reactor but over
shorter periods of times, e.g. a few hours or days, in the hope that
the observed (accelerated) material degradation can be used as a predictor
of the performance of the same material during its lifetime in a real
reactor. The premise of accelerated materials testing is that materials
theory and numerical simulations can provide a reliable connection
between the accelerated tests and the material lifetime performance
predictions. To serve this purpose, material simulations should meet
two conditions. First, accurate material models need to be developed
and validated against experimental measurements. Second, the simulation
algorithms need to be efficient to enable computational predictions
of materials performance under reactor conditions. The performance
data presented in Table \ref{FPKMC_performance} suggests that FPKMC
can meet this second challenge. Here we continue to focus on the thin-film
model studied in Section \ref{SectionRadiationDamage} and use our
FPKMC code to simulate damage accumulation at two different dose rates,
a \emph{high} dose rate of $1.5\cdot10^{-4}$dpa/s typical of of accelerated
experiments in material testing facilities such as JANNUS \citet{JANNUS},
and a \emph{low} dose rate of $1.5\cdot10^{-8}$dpa/s typical of the
existing nuclear reactors. We ran both simulations to a total dose
of $10$ dpa, which required several CPU days per sample on a common
workstation.

Direct comparison of two simulations performed at the same temperature
$T=200^{o}C$ revealed very different kinetics and end-of-dose damage,
which is not surprising given that much more time is available for
damage annealing (healing) at the slow (reactor) time scales. In order
to enable scaling from high dose rates to low dose rates, it was proposed
\citet{AcceleratedTesting_Scaling} to raise the temperature in the
high dose rate irradiation test so as to preserve the ratio of the
damage insertion rate to the rate of defect diffusion. Such scaling
would be exact if there were only one evolution mechanism whose rate
can be adjusted by changing the temperature. However, even in the
simple model of $\alpha$-iron considered here, there is a whole spectrum
of mechanisms and associated rates with different temperature activation
parameters. One can only hope that an approximate scaling can be achieved
by adjusting the rate of just one dominant mechanism that controls
the overall rate of damage accumulation. In an extensive series of
numerical experiments we observed that, within the simple model considered
here, the overall rate and character of damage evolution appears to
be controlled by the ratio of vacancy diffusion to the irradiation
dose rate. This is likely because, at all temperatures of interest
here, the interstitials are much more mobile than the vacancies and
disappear nearly instantly following the insertion of a Frenkel pair,
leaving the less mobile vacancies to diffuse and cluster in the absence
of interstitials.

\begin{figure}
\begin{centering}
\includegraphics[width=0.95\textwidth]{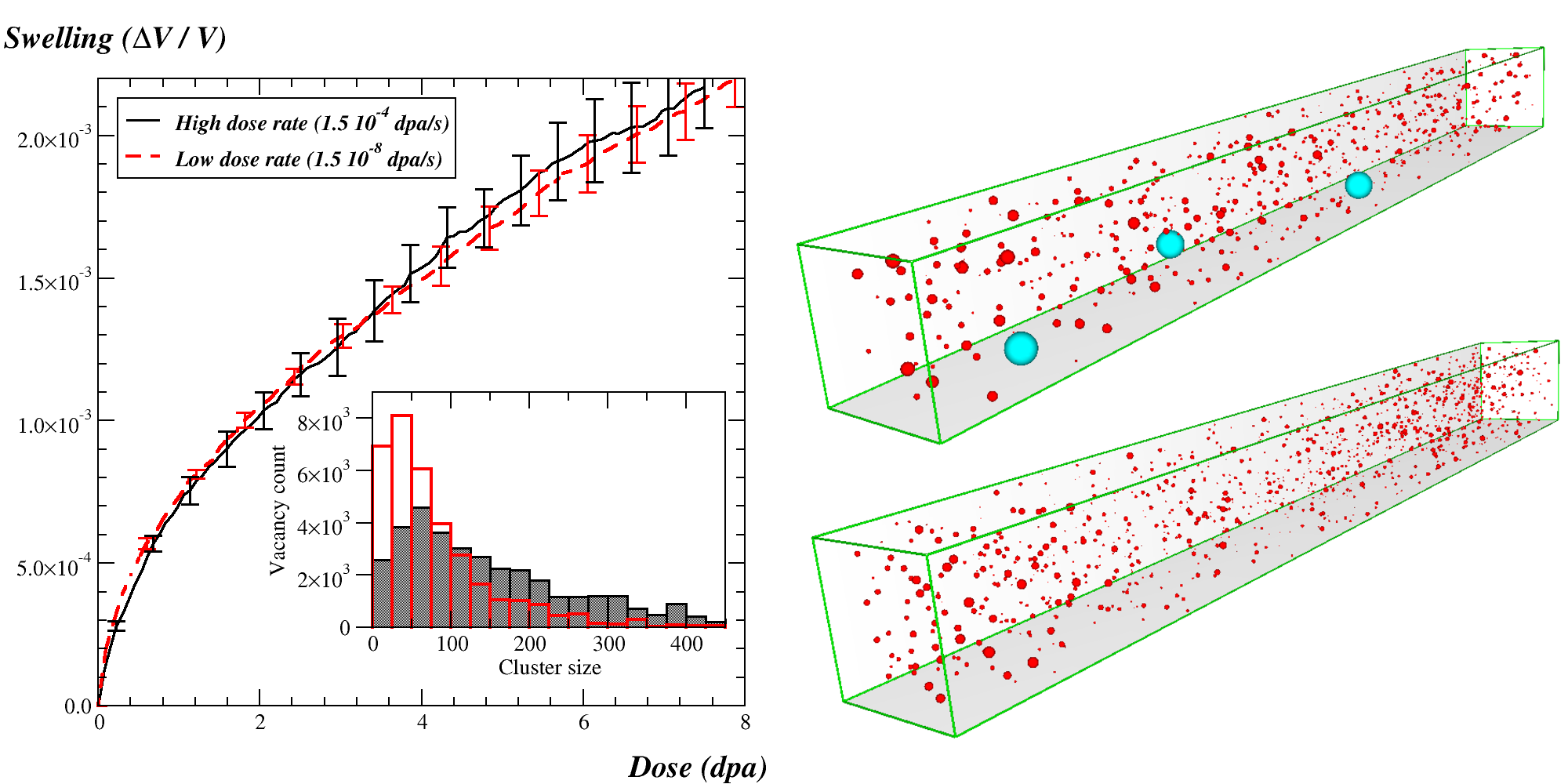}
\par\end{centering}

\caption{\label{Accelerated.comparison.10dpa}State of damage in a thin $\alpha-$Fe
film irradiated by electrons to the total dose of 10 dpa:\protect \\
(Top right) At the high dose rate of $1.5\cdot10^{-4}$dpa/s and $T=262\celsius$
and (Bottom right) At the low dose rate of $1.5\cdot10^{-8}$dpa/s
and $T=130\celsius$.\protect \\
(Left) Volume fraction of vacancies (swelling) as a function of damage
dose. The vacancy fraction was obtained by counting together all vacancies
in the simulation volume, both in the vacancy monomers or vacancy
clusters, and dividing the sum by the total number of atomic sites.
The black solid curve is the swelling kinetics under the high dose
rate/high temperature and the dashed red curve is the same kinetics
under the low dose rate/low temperature irradiation conditions. Both
curves were obtained by averaging over 10 independent simulation runs
for each irradiation condition, and the vertical bars show the estimated
statistical errors. The inset in the bottom figure shows a histogram
of the distribution of vacancy cluster sizes at a dose of 5 dpa for
both the high dose rate/high temperature irradiation (shaded gray
bars) and for the low dose rate/low temperature irradiation (red bars).}

\end{figure}

Figure \ref{Accelerated.comparison.10dpa} compares the state of damage
reached at $T=262\celsius$ and high dose rate, with that reached
at $T=130\celsius$ at low dose rate. The kinetics of damage accumulation
and the resulting populations of voids are similar to each other but
the scaling is only approximate and some differences in the resulting
microstructures are noticeable. Most visibly, a few large interstitial
clusters form at the high dose rate where the time interval between
successive Frenkel-pair insertions is comparable to the lifetime of
inserted interstitials. On the contrary, at the lower dose rate the
newly inserted interstitials disappear (at the free surfaces or through
annihilation with the vacancies) well before the next Frenkel pair
is inserted. Additionally, as seen from the histograms shown in the
insert, the average size of the vacancy clusters is larger at the
higher dose rate. The approximate agreement in the overall swelling
supports the idea that it may be possible to compensate the enhanced
rate of damage accumulation in accelerated tests by raising the test
temperature so that the resulting damage is approximately the same
as in the reactor but at a lower temperature. That the scaling is
only approximate is less than surprising considering that multiple
rate processes, each with its own temperature dependence, act together
to produce the resulting damage kinetics.

\section{\label{SectionConclusions}Conclusions}

We have described an asynchronous event-driven algorithm for diffusion
kinetic Monte Carlo simulations of diffusion-reaction particle systems,
based on the First Passage Kinetic Monte Carlo method first proposed
in Ref. \citet{FPKMC_PRL} and described in more detail in Ref. \citet{FPKMC1}.
The First Passage Kinetic Monte Carlo (FPKMC) algorithm avoids long
sequences of small diffusive hops commonly used to bring particles
to collisions, by enabling large super-hops sampled from exact (semi)analytical
solutions for diffusion Green's functions (propagators) in spatially
isolated protective regions each containing just one or two particles.
The FPKMC algorithm is exact to the extent that the system's stochastic
trajectory is sampled from the exact diffusion-reaction Master Equation
for the $N$-particle system, to within the accuracy of the single-
and two-body Green's functions. At the same time, for a range of simulations
reported here and in Ref. \citet{FPKMC1}, the new algorithm is several
orders of magnitude more efficient than the existing (approximate)
hopping-based algorithms.

In this work we extended the FPKMC method to considerably more complex
simulations in which particle diffusion is just one among many competing
processes. We described generalizations and algorithmic components
necessary to handle systems with multiple particle species and multiple
reactions, including annihilation, clustering, emission, and particle
birth and death. We focused on the case of hard spheres that are more
appropriate than cubes for a multitude of intended applications. Handling
hard sphere collisions as a first-passage process turned out to be
more complicated than the case of cubes which necessitated a hopping-based
approximation for the pair propagators. The resulting solution to
the sphere collision problem was found practical and efficient but
raised a more general issue of combining the asynchronous event-driven
framework employed in FPKMC with the more traditional synchronous
time-driven simulations. Our successful implementation of time-driven
pair propagators in FPKMC points to hybrid time-driven/event-driven
algorithms as a promising direction for future research.

The new implementation of the FPKMC algorithm has proven suitable
for simulations of damage accumulation in materials subjected to irradiation
by high-energy particles. The accuracy of the FPKMC algorithm and
our implementation was validated on several test problems by comparison
to traditional (object) KMC algorithms developed for diffusion-reaction
simulations in the continuum (BIGMAC) and on the lattice (LAKIMOCA).
The new algorithm is shown to perform well in a wide range of radiation
conditions enabling, for the first time, simulations of irradiated
materials to large technologically relevant radiation doses (e.g.,
$10\mbox{dpa}$) on a serial workstation. Closing the gap between
the short time horizon of traditional KMC simulations and long material
life in the reactor required to gain several orders of magnitude in
computational efficiency. In FPKMC, this gain is achieved entirely
through an exact factorization of the difficult $N$-body reaction-diffusion
problem into one- and two-body problems. 

With its efficiency, the new FPKMC method can make a significant impact
on the important area of accelerated material testing for next-generation
nuclear reactor designs. There is a strong synergy between atomistic
KMC simulations and experiments carried out in accelerated testing
facilities. The accuracy of an atomistic KMC model can be improved
by expanding its mechanism catalog and obtaining more accurate values
of model parameters. For this purpose, computationally efficient KMC
simulations can be used to explore and identify experimental conditions
in which accelerated material tests would be most informative for
model validation, e.g. most sensitive to a particular mechanism or
model parameter. Furthermore, the same simulations can be used to
fine-tune the KMC models for conditions typical of real reactors.
The approximate scaling predicted in our FPKMC simulations provides
exactly the right kind of connection between simulations and experiments.
Once the accuracy of the material models is established, FPKMC simulations
can be used to extrapolate from accelerated material tests into relevant
but inaccessible conditions of nuclear reactors without relying on
any approximate scaling. To quantify the reliability of such computational
extrapolations, FPKMC simulations can be used to assess the uncertainties
in computational predictions of accumulated damage given the uncertainties
in model parameters, similarly to what is routinely done in climate
modeling. Ultimately, efficient KMC simulations can and should become
an integral component of reliable material testing programs.

In addition to allowing reactor material simulations on technologically-relevant
time-scales, the new method may prove enabling in other areas of science,
in engineering and in finance. One particularly attractive application
for the new method is in cell biology, where multiple reaction-diffusion
mechanisms conspire to produce a wide variety of specific biological
responses. Parallelization of the FPKMC algorithm could further extend
length and time scales accessible to diffusion-reaction simulations,
even though the asynchronous event-driven algorithms are notoriously
difficult to parallelize \citet{AED_Review}.

\section{Acknowledgments}

We would like to thank Christophe Domain, Pär Olsson, and Charlotte
Becquart for helpful discussions and sharing with us their OKMC simulation
data. We would also like to thank Mihai-Cosmin Marinica for testing
our FPKMC code extensively and Maria-Jose Caturla for sharing with
us her library of collision cascades. This work was performed under
the auspices of the U. S. Department of Energy by Lawrence Livermore
National Laboratory under Contract DE-AC52-07NA27344. This work was
supported by the Office of Laboratory Directed Research at LLNL and
the Office of Basic Energy Sciences U. S. Department of Energy. 

\begin{appendix}

\section{\label{AppendixMonomer}Sampling the Single-Sphere Propagators}

The solution to the diffusion problem inside a sphere with absorbing
boundaries $c(r,t)$ can only be represented in closed-form in the
Laplace domain and thus one must resort to rapidly-converging series
solutions, as given in Eqs. (\ref{c_mono_short}) and (\ref{c_mono_long})
for short and long times, respectively. As explained in more detail
in Ref. \citet{FPKMC1}, efficient rejection sampling of such distributions
can be performed by truncating the series if the truncated series
can be augmented with an upper and lower bound on the true distribution,
tighter and tighter as more terms are added to the series. For example,
estimates of the absolute value of the remainder in the truncated
series can be used to provide bounds.

Randomly sampled points can be rejected if they are above the upper
bound or accepted if they are below the lower bound, or, if neither,
the next term added to the series to compute tighter bounds. With
rapidly converging series and tight bound estimates, typically only
a few of the leading terms in the series need to be computed while
still providing \emph{exact} sampling of the target distribution.
In Ref. \citet{FPKMC1}, series for the various bounds were given
for the case of a point particle diffusing with diffusion coefficient
$D=1$ inside a one dimensional interval of length one (starting from
the center). Here we give the corresponding three dimensional results,
i.e., the case of a point diffusing inside a unit three dimensional
sphere (starting from the center).

\subsubsection{First-passage propagator}

Integration of $c(r,t)$ over the unit sphere yields the survival
probability in two infinite series forms, a short time series

\begin{equation}
S(t)=S_{n\rightarrow\infty}(t),\end{equation}
where\[
S_{n}=(2\pi t^{3})^{-\frac{1}{2}}\sum_{m=-n-1}^{n}\left[2m+\frac{(1+2m)^{2}}{2t}\right]\exp\left[-\frac{(1+2m)^{2}}{4t}\right],\]
and a long time series

\begin{equation}
S(t)=-2\pi^{2}\sum_{m=1}^{\infty}(-1)^{m}m^{2}e^{-m^{2}\pi^{2}t}.\end{equation}

We select a switchover time $\tau$ between the short- and long-time
series such that $1/\pi^{2}\leq\tau\leq1/4$ and use the piecewise
smooth function $C(t)$ as an over-estimator for the survival probability
at all times,\[
C(t)=\left\{ \begin{array}{c}
\frac{S(\tau)}{S_{s}(\tau)}S_{s}(t)\mbox{ for }t<\tau\\
S_{l}(t)\mbox{ for }t>\tau,\end{array}\right.\]
where $S_{s}(t)=(\frac{1}{t}-2)e^{-\frac{1}{4t}}/\sqrt{2\pi t^{3}}$
and $S_{l}(t)=2\pi^{2}e^{-\pi^{2}t}$ are the leading terms of the
short- and long-time series for the survival probability. A sample
exit time $t$ is obtained by solving $C(t)=\xi$, where $\xi$ is
random number uniformly distributed in $[0,1)$. Solving this equation
for $t<\tau$ can be done efficiently by Newton iteration. To obtain
the sequence of converging lower and upper bounds necessary for rejection
sampling, note that the terms of the long-time series alternate in
sign and decrease in magnitude with increasing $m$ and can themselves
serve as the needed bounds. For the short-time series we can bound
the remainder $R_{n}(t)=S(t)-S_{n}(t)$ with

\[
0\leq R_{n}(t)\leq\left(4\pi t^{3}\right)^{-\frac{1}{2}}\left[\frac{(1+2m)^{2}}{t}-1+2m\right]\exp\left[-\frac{(1+2m)^{2}}{4t}\right].\]
In our implementation we use $\tau=0.243$, which gives a rejection
ratio of about $0.6$\%. The maximum relative error in the survival
probability is $C(t)/S(t)-1\thickapprox7\cdot10^{-3}$, so that $C(t)$
is an acceptable approximation for $S(t)$ without rejection sampling.

\subsubsection{No-passage propagator}

To enable efficient rejection sampling, we need a tight over-estimator
$C(r;t)$ for the no-passage probability distribution $c_{1}(r;t)$
at arbitrary time $t$. Here we construct such a function by stitching
together two different expressions appropriate for times shorter and
longer than a switchover time $\tau$ ($1/\pi^{2}\leq\tau\leq1/4$).
For times $t<\tau$ the leading term ($m=0$) in the short-time series
solution (1) is a reasonable over-estimator:

\[
C_{s}(r;t)=(4\pi t)^{-\frac{3}{2}}e^{-\frac{r^{2}}{4t}}.\]
On the other hand, for times $t>\tau$ a good over-estimator is given
by

\[
C_{l}(r;t)=\frac{\sin\pi r}{2r}\left[e^{-\pi^{2}t}+\frac{1}{\pi^{2}t}\left(1+\frac{1}{4\pi^{2}t}\right)e^{-4\pi^{2}t}\right].\]

Sampling from $C_{s}(r;t)$ entails evaluation of one inverse error
function while sampling from $C_{l}(r;t)$ requires solving for $r$
the following equation

\[
\frac{1}{\pi}\sin\pi r-r\cos\pi r=\xi,\]
where $\xi$ is a random number uniformly distributed on $[0,1)$.
The solution can be found by Newton iteration using $r=\sqrt[3]{3\xi/\pi^{2}}$
as a starting guess if $\xi\leq1/2$ and $r=1-\frac{1}{\pi}\sqrt{2(1-\xi)}$
if $\xi>1/2$.

The same two series solutions can be used to obtain a sequence of
increasingly tight bounds on the no-passage distribution. For short
times, we approximate $c(r,t)$ with the partial sum

\[
c_{m}(r,t)=\sum_{k=-m}^{m}(1+\frac{2k}{r})\exp\left[-\frac{(r+2m)^{2}}{4t}\right],\]
and use the following lower and upper bounds for the remainder term,
$R_{l}\leq c(r,t)-c_{m}(r,t)\leq R_{u}$,

\begin{eqnarray*}
R_{l} & = & \left(1-2\frac{m+1+t}{r}\right)\exp\left[-\frac{(r-2m-2)^{2}}{4t}\right]+\frac{2t}{r}\exp\left[-\frac{(r+2m+2)}{4t}\right]\\
R_{u} & = & \left(1+2\frac{m+1+t}{r}\right)\exp\left[-\frac{(r+2m+2)^{2}}{4t}\right]+\frac{2t}{r}\exp\left[-\frac{(r-2m-2)}{4t}\right],\end{eqnarray*}
For long times, tight lower and upper bounds on the no-passage probability
distribution are obtained from the following bound for the magnitude
of the remainder

\[
\left|4\pi r^{2}\sum_{k=m+1}^{\infty}\frac{k}{2r}\sin(k\pi r)e^{-m^{2}\pi^{2}t}\right|\leq\left[2\pi r(m+1)+\frac{r}{\pi t}\right]e^{-(m+1)^{2}\pi^{2}t}.\]

\section{\label{SectionNNL}Near-Neighbor List (NNL) Method}

The \emph{near-neighbor list} (NNL) method \citet{Event_Driven_HE}
is a neighbor search technique that is superior to the linked list
cell method in conditions where particles do not change neighbors
over many events. The essential idea is to enclose every protective
region $\mathcal{P}_{i}$ inside a \emph{bounding neighborhood} $\mathcal{N}_{i}$,
$\mathcal{P}_{i}\subset\mathcal{N}_{i}$. This bounding neighborhood
is fixed while the particle and its protection change as the particle
moves around, until the particle comes close to the boundary of $\mathcal{N}_{i}$
at which point a new $\mathcal{N}_{i}$ is constructed. In principle,
one can treat the boundary of $\mathcal{N}_{i}$ as a first-passage
surface, however we simply rebuild the bounding neighborhood whenever
the particle comes close to its boundary.

The linked list $\mbox{NNL}(i)$ lists all other neighborhoods that
intersect neighborhood $\mathcal{N}_{i}$ (hard walls or other boundaries
may also be near neighbors). This is used to identify potential \emph{interactions}
of particle $i$ and can be reused until the particle core $\mathcal{C}_{i}$
approaches the boundary of $\mathcal{N}_{i}$. This results in significant
savings of computational effort if particle motion is localized and
the particles experience numerous displacements before leaving their
bounding neighborhood. Note that the LLC method is still used to build
$\mathcal{N}_{i}$ and $\mbox{NNL}(i)$ which keeps the maximal cost
of pairwise searches at $O(N)$ instead of $O(N^{2})$. In our implementation
$\mathcal{N}_{i}$ is build as a sphere concentric with the particle
and with the diameter larger than $2\mu R_{i}$ but smaller than the
cell size, where $\mu>1$ is a parameter. 

Even the NNL method can become inefficient when some particles are
much larger than others, e.g. large clusters formed by coalescence
of defects in radiation damage modeling. In such cases, if the cells
are still maintained larger than the size of the largest particle,
the same cells may contain many small particles making the search
for near neighbors expensive. On the other hand, if the cells are
kept reasonably small, neighbor searches need to examine many small
cells in order to account for all near neighbors of the large particles.
The idea of the \emph{bounding sphere complexes} (BSCs) \citet{Event_Driven_HE}
method is to use small cells but to cover $\mathcal{N}_{i}$ with
a collection of $N_{BSC}$ spheres, each smaller than the cell size.
The so-constructed sphere complex remains immobile until $\mathcal{N}_{i}$
changes, which would occur infrequently if the large particles move
slowly or are immobile. The small spheres forming BSCs are inserted
in the LLCs and near neighbors of each large particle are found by
searching for the near neighbors of each constituent small sphere
in the corresponding BCS.

\end{appendix}


\end{document}